\documentclass[12pt]{article}
\pdfoutput=1
\usepackage{color}
\usepackage{amssymb,amsmath,bm}
\usepackage{epsf}
\usepackage{epsfig}
\usepackage{afterpage}
\usepackage{longtable}
\usepackage{caption}
\usepackage{cite}
\usepackage{latexsym,mathrsfs}
\usepackage{graphics}
\usepackage{url}
\usepackage{paralist}
\usepackage{subcaption}
\usepackage{float}

\usepackage[dvipsnames]{xcolor}
\usepackage[linktoc=page,bookmarks=false,colorlinks=false,linkbordercolor=RoyalBlue,citebordercolor=ForestGreen,urlbordercolor=CornflowerBlue]{hyperref}

\usepackage[utf8x]{inputenc}

\usepackage{booktabs}

\newcommand{\ord}{\mathcal{O}}

\newcommand{\IM}{\rm{Im}}
\newcommand{\RE}{\rm{Re}}

\newcommand{\tev}{\, {\rm TeV}}
\newcommand{\gev}{\, {\rm GeV}}
\newcommand{\mev}{\, {\rm MeV}}

\newcommand{\vcb}{|V_{cb}|}

\newcommand{\vub}{|V_{ub}|}

\newcommand{\kepe}{\kappa_{\varepsilon^\prime}}
\newcommand{\keps}{\kappa_{\varepsilon}}

\def\epe{\varepsilon'/\varepsilon}
\newcommand{\beq}{\begin{equation}}
\newcommand{\eeq}{\end{equation}}
\newcommand{\be}{\begin{equation}}
\newcommand{\ee}{\end{equation}}
\newcommand{\bi}{\begin{itemize}}
\newcommand{\ei}{\end{itemize}}
\newcommand{\ba}{\begin{array}}
\newcommand{\ea}{\end{array}}
\newcommand{\beqa}{\begin{eqnarray}}
\newcommand{\eeqa}{\end{eqnarray}}
\newcommand{\bea}{\begin{eqnarray}}
\newcommand{\eea}{\end{eqnarray}}
\newcommand{\beqn}{\begin{eqnarray}}
\newcommand{\eeqn}{\end{eqnarray}}

\newcommand{\eps}{\epsilon}

\definecolor{red}{cmyk}{0,1,1,0.4}

\def\kpn{K^+\rightarrow\pi^+\nu\bar\nu}
\def\klpn{K_{L}\rightarrow\pi^0\nu\bar\nu}

\newcommand{\wc}[3][{}]{\big[{\cal C}_{#2}^{#1}\big]_{#3}}

\newcommand{\Wc}[2][{}]{{\cal C}_{#2}^{#1}}

\newcommand{\op}[3][{}]{[{\cal O}_{#2}^{#1}]_{#3}}

\newcommand{\Op}[2][{}]{{\cal O}_{#2}^{#1}}

\usepackage{fancyhdr}
\advance \headheight by 3.0truept       

\begin{document}

\begin{flushright}
    {AJB-20-3\\
    UdeM-GPP-TH-19-280}\\

\end{flushright}

\medskip

\begin{center}
{\LARGE\bf
\boldmath{Another SMEFT Story: $Z^\prime$ Facing New Results on $\epe$, $\Delta M_K$ and $K\to\pi\nu\bar\nu$}}
\\[0.8 cm]
{\bf Jason~Aebischer$^{a}$,  Andrzej~J.~Buras$^{b}$ and
Jacky Kumar$^{c}$
 \\[0.5 cm]}
{\small
$^a$ Department of Physics, University of California at San Diego, La Jolla, CA 92093, USA\\
$^b$TUM Institute for Advanced Study, Lichtenbergstr. 2a, D-85747 Garching, Germany\\
$^c$Physique des Particules, Universite de Montreal, C.P. 6128, succ.  centre-ville,\\
Montreal, QC, Canada H3C 3J7}
\\[0.5 cm]
\footnotesize
E-Mail:
\texttt{jaebischer@physics.ucsd.edu},
\texttt{andrzej.buras@tum.de},
\texttt{jacky.kumar@umontreal.ca}
\\[0.2 cm]
\end{center}

\vskip0.41cm

\begin{abstract}
\noindent
Recently the RBC-UKQCD lattice QCD collaboration presented new results for the hadronic matrix elements relevant for the ratio $\epe$ in the
Standard Model (SM) albeit with significant uncertainties. With the present knowledge of the Wilson coefficients and isospin breaking effects there is still
a {sizable} room left for new physics (NP) contributions to $\epe$ which could both {\em enhance or suppress}  this ratio to agree with the data.
The  new SM
value for the $K^0-\bar K^0$ mass difference $\Delta M_K$ from RBC-UKQCD  is
on the other hand by $2\sigma$
above the data hinting for NP required to {\em suppress}  $\Delta M_K$.
Simultaneously the most recent results for $\kpn$ from NA62 and for $\klpn$ from KOTO still allow for significant NP contributions. {We point out
  that the suppression of $\Delta M_K$ by NP requires the presence of new
  CP-violating phases with interesting implications for $K\to\pi\nu\bar\nu$,
  $K_S\to\mu^+\mu^-$ and $K_L\to\pi^0\ell^+\ell^-$ decays.
  Considering a $Z^\prime$-scenario within the SMEFT we analyze the
  dependence of all these observables on the size  of NP still allowed
  by the data on $\epe$. {The hinted $\Delta M_K$ anomaly together
    with the $\varepsilon_K$ constraint implies  in the presence
    of only left-handed (LH) or right-handed (RH) flavour-violating $Z^\prime$ couplings
    strict correlation between   $\kpn$ and $\klpn$ branching ratios so
    that they are either simultaneously enhanced
    or suppressed relative to SM predictions. An anticorrelation can only be obtained in the presence of both LH and RH couplings.}
Interestingly, the NP QCD penguin scenario for $\epe$ is excluded by SMEFT renormalization group effects in $\varepsilon_K$ so that NP effects in $\epe$
are governed by electroweak penguins.}
  We also
investigate for the first time whether  the presence of a heavy $Z^\prime$ with flavour violating couplings could generate through top Yukawa renormalization group effects  FCNCs mediated by the SM $Z$-boson. The outcome turns out to be very interesting.
\end{abstract}

\setcounter{page}{0}
\thispagestyle{empty}
\newpage

\setcounter{tocdepth}{2}
\tableofcontents

\newpage

\section{Introduction}
The ratio
$\epe$ that  measures the size of direct CP violation in $K_L\to\pi\pi$ decays
relative to the indirect CP violation described by $\varepsilon_K$ and the
rare decays $\kpn$ and $\klpn$ have been already for many years together with the $\Delta I=1/2$
rule, $K_{L,S}\to \mu^+\mu^-$ and $K_L\to \pi^0\ell^+\ell^-$ decays the stars of
Kaon flavour physics \cite{Buras:2013ooa}. The $K_L$--$K_S$ mass difference
$\Delta M_K$ remained due to large theoretical uncertainties  until recently under the shadow of these decays
although it played a very important role in the past in estimating successfully  the charm quark mass prior to its discovery \cite{Gaillard:1974hs}. However, recently progress in evaluating $\Delta M_K$
within the SM has been made by the RBC-UKQCD  collaboration \cite{Bai:2014cva,Christ:2015pwa,Bai:2018mdv} so that  $\Delta M_K$
begins to play again an important role in phenomenology, not only to bound
effects of NP contributions \cite{Gerard:1984bg,Gabbiani:1996hi,Bona:2007vi,Isidori:2010kg,Silvestrini:2018dos,Calibbi:2019lvs}, but also to help {identify} what this NP
could be. But as stressed in \cite{Buras:2013ooa} and in particular in
 \cite{Buras:2015jaq} such an identification is only possible by considering all the stars
of Kaon physics simultaneously and also invoking observables from other meson systems.

{The RBC-UKQCD lattice QCD collaboration presented very recently
  new results for the hadronic matrix elements relevant for the ratio $\epe$.
  Using the Wilson coefficients at the NLO level and {not including isospin} breaking and NNLO QCD effects they  find \cite{Abbott:2020hxn}
\begin{align}
  \label{RBCUKQCD}
  (\epe)_{\rm SM} &
  = (21.7 \pm 8.4) \times 10^{-4} \,,
\end{align}
where statistical, parametric and systematic uncertainties have been added in
quadrature.

However, as already demonstrated in \cite{Aebischer:2019mtr}, the inclusion
of the effects in question, that are absent in (\ref{RBCUKQCD}) is important.
Including the isospin breaking contributions, recently calculated in
\cite{Buras:2020pjp}
 and the NNLO QCD corrections to electroweak penguin contributions \cite{Buras:1999st},
the result in (\ref{RBCUKQCD})
is changed to \cite{Buras:2020pjp,Aebischer:2020jto}\footnote{Without the presence of $\eta-\eta^\prime$ mixing in the estimate of isospin-breaking corrections, as done in \cite{Cirigliano:2019cpi}, one would find instead $ (\epe)_{\rm SM} = (17.4 \pm 6.1) \times 10^{-4}$ \cite{Buras:2020pjp,Aebischer:2020jto}.}
\begin{align}
  \label{ABB}
  (\epe)_{\rm SM}  & = (13.9 \pm 5.2) \times 10^{-4} \,,
\end{align}
which compared with the
experimental world average from NA48 \cite{Batley:2002gn} and
  KTeV \cite{AlaviHarati:2002ye, Worcester:2009qt} collaborations,
  \begin{align}
    \label{EXP}
    (\epe)_\text{exp} &
    = (16.6 \pm 2.3) \times 10^{-4} \,,
  \end{align}
  shows a very good agreement of the SM with the data, albeit leaving still
   much room for NP contributions. Presently values as low as 5 or as high
  as 25 in these units cannot be excluded.

  While this result allows for both positive and negative  NP contributions
  to  $\epe$ to agree with the data, the new SM
value for the $K^0-\bar K^0$ mass difference $\Delta M_K$ from  RBC-UKQCD \cite{Bai:2018mdv}
\be\label{NPDMK}
(\Delta M_K)_\text{SM}=7.7(2.1) \times 10^{-15} \gev,\qquad (\Delta M_K)_\text{exp} = 3.484(6) \times 10^{-15} \gev\,,
\ee
hints at the  $2\sigma$ level at the presence of NP
required to {\em suppress}  $\Delta M_K$ relative to its SM value.

As noted already in \cite{Buras:2015jaq} the suppression of $\Delta M_K$  is only possible in the presence
of new CP-violating couplings. This could appear surprising at first sight,
{since} $\Delta M_K$ is a CP-conserving quantity but simply follows from the
fact that the BSM shift $(\Delta M_K)_\text{BSM}$ is proportional to the square of a complex
$g_{sd}$ coupling so that
\be
(\Delta M_K)_\text{BSM}= c~\RE[g_{sd}^2]= c\left[(\RE[ g_{sd}])^2-(\IM[ g_{sd}])^2\right], \qquad c>0.
\ee
The required negative contribution implies automatically NP contributions
to $\epe$ and also to rare decays  $K\to\pi\nu\bar\nu$,
$K_S\to\mu^+\mu^-$ and $K_L\to\pi^0\ell^+\ell^-$, provided this NP involves
non-vanishing flavour conserving $q\bar q$ couplings in the case of $\epe$
and non-vanishing $\nu\bar\nu$ and $\mu^+\mu^-$ couplings in the case
of {the} rare $K$ decays in question.

Now,  as pointed out in an important paper by Monika Blanke
eleven years ago \cite{Blanke:2009pq},  in the presence of a
strict correlation between NP contributions to $\Delta S=1$ and $\Delta S=2$
processes and assuming no  significant NP contributions to $\varepsilon_K$
implies two narrow branches in the  $(\kpn,\klpn)$ plane , namely
 \begin{itemize}
 \item
   a branch parallel to the Grossman-Nir {(GN)} bound \cite{Grossman:1997sk} on
 which both branching ratios can either simultaneously increase or decrease
 relative to SM values,
\item
  a horizontal narrow branch on which there is no NP contribution to $\klpn$ because
  of the absence of flavour-violating complex couplings.
 \end{itemize}

 This is in particular the case of NP entering already at tree-level
 with only  left-handed or right-handed flavour-violating NP  couplings with
 the prominent example of $Z^\prime$ models in which the $Z^\prime \bar s d$ coupling enters both $\kpn$ and $\klpn$ as well as $\varepsilon_K$.

  The hinted anomaly in $\Delta M_K$, requiring the imaginary couplings
   to be present, excludes the horizontal branch so
   that the full action of NP in this case  happens only on the second branch to be called  {\it MB-branch} in what follows.

   But in \cite{Blanke:2009pq} {a} possible impact of $\Delta M_K$ has not been discussed. {Therefore, under the assumption of} significant NP contributions to $\varepsilon_K$ but still considering only scenarios with left-handed or right-handed flavour-violating $Z^\prime$ couplings leads to significantly broader  branches than when $\varepsilon_K$ from the
   SM agreed with the data. However, as we will demonstrate in the present paper,
   the removal of the  $\Delta M_K$ anomaly combined with renormalization
   group Yukawa top effects implies still a rather narrow MB-branch.

 In this context it is interesting to observe that the most recent
result for $\kpn$ from NA62 \cite{NA62new} and the  $90\%$ confidence level (CL) upper bound on $\klpn$  from KOTO  \cite{Ahn:2018mvc} read respectively
\be\label{EXP19}
\mathcal{B}(\kpn)_\text{exp}=(4.7^{+7.2}_{-4.7})\times 10^{-11}\,,\qquad
\mathcal{B}(\klpn)_\text{exp}\le 3.0\times 10^{-9}\,,
\ee
to be compared with the SM predictions \cite{Buras:2015qea,Bobeth:2016llm}
\be\label{KSM}
\mathcal{B}(\kpn)_\text{SM}= (8.5^{+1.0}_{-1.2})\times 10^{-11}\,,\qquad
\mathcal{B}(\klpn)_\text{SM}=(3.2^{+1.1}_{-0.7})\times 10^{-11}\,.
\ee

In their most recent status report \cite{KOTOnew} on $\klpn$ the KOTO collaboration presented data on four candidate events in the signal region, finding
\be\label{KOTO}
\mathcal{B}(\klpn)_\text{KOTO}=2.1^{+2.0(+4.1)}_{-1.1(-1.7)}\times 10^{-9}\,,
\ee
at the 68 (95) \% CL. The central  value is by a factor of 65
  above the central SM prediction and in fact violates the GN bound
  which at the $90\%$ CL together with the present NA62 result for $\kpn$ amounts
  to $0.8\times 10^{-9}$. Theoretical analyses of this interesting data
  can be found in \cite{Kitahara:2019lws,He:2020jly,He:2020jzn,Fuyuto:2014cya}.

Evidently there is still much room for NP left in these decays. In particular, a pattern  in which $\kpn$ is {\em suppressed} and
$\klpn$ is {\em enhanced} by NP is hinted by the new data. As pointed out
already in \cite{Blanke:2009pq} and seen in the plots in \cite{Blanke:2009pq,Buras:2015yca} this pattern is only possible in the presence of both
left-handed and right-handed flavour-violating $Z^\prime$  couplings to quarks which with moderate fine-tuning allows to avoid the constraint from
$\varepsilon_K$, so that regions in the  $(\kpn,\klpn)$ plane outside the MB-Branch are possible. We will return to this issue in Section~\ref{LRSCENARIO},
  but we stress already here, following \cite{Blanke:2009pq},
  that generally in NP scenarios in which NP contributions to $\Delta S=1$ and $\Delta S=2$ are not related to each other, different oases in the  $(\kpn,\klpn)$ plane  outside the MB-Branch could be
  occupied.  As evident from the plots in \cite{Buras:2001af,Buras:2015yca} the simplest example are models with minimal flavour violation (MFV).
   There the   correlation between $\kpn$ and $\klpn$
   results from the same real valued loop function
  $X$ entering these two processes. This function
   is a priori unrelated to NP contributions in $\Delta S=2$ processes and therefore $\Delta S=2$ constraints are avoided. On the other hand in the absence
   of new complex flavour-violating phases in MFV models the suppression
of $\Delta M_K$ is not possible. This is reminiscent of lower bounds on $\Delta M_{s,d}$ present in these models \cite{Blanke:2006yh,Blanke:2016bhf}.

It has been pointed out already in  \cite{Buras:2015jaq}
that various patterns of NP in rare $K$ decays in correlation with NP in $\epe$   can naturally be realized
in models with tree-level FCNCs mediated by a heavy $Z^\prime$ with masses still
in the reach of ATLAS and CMS but also for higher masses. But whereas
in \cite{Buras:2015jaq} the scenarios with enhanced $\kpn$ and $\klpn$
have been primarily considered,  a novel pattern in which $\kpn$ is {\em suppressed} and
$\klpn$ is {\em enhanced} by NP  possibly hinted by the new data has not been considered
there.

With the new information from RBC-UKQCD on $\epe$ and $\Delta M_K$, the new analyses of $\epe$ in \cite{Buras:2020pjp,Aebischer:2020jto}
and the new data from NA62 and KOTO, it is of interest to ask how the
$Z^\prime$ scenarios considered in  \cite{Buras:2015jaq} and the new ones
face the new developments listed above.

The goal of the present paper is to answer this question, but our paper should not be
considered as the numerical update of the analysis in \cite{Buras:2015jaq} motivated by
the new input from RBC-UKQCD, NA62 and KOTO collaborations. The
reason is that
in contrast to  \cite{Buras:2015jaq}, that included only QCD renormalization group effects, we will perform  a complete SMEFT analysis, that takes in particular  into
account important top Yukawa effects, which modify significantly the properties of a $Z^\prime$ responsible for the pattern of NP effects in question.
In
particular we point out that the so-called QCD penguin scenario for $\epe$,  considered in \cite{Buras:2015jaq},
in which at { the} NP scale only QCD penguin operators have non-vanishing Wilson coefficients,
is excluded due to Yukawa renormalization group effects { on $\varepsilon_K$}
when NP contributions to $\epe$ and $\Delta M_K$ are considered simultaneously.
We demonstrate this effect both analytically and numerically.

In models with vector-like quarks the operators $\psi^2 H^2D$, listed in Table~\ref{tab:psi2H2D}, are generated at the matching scale, implying FCNCs mediated by the
SM $Z$-boson. They can be enhanced through RG Yukawa top quark effects
with an important impact on the phenomenology \cite{Bobeth:2016llm,Bobeth:2017xry,Endo:2018gdn}. Such
operators have vanishing Wilson coefficients in $Z^\prime$  models at tree-level
{if the $(H^\dagger i D_{\mu} H) Z_{\mu}^\prime$ coupling is set to zero.
However, they} are generated again through RG Yukawa top quark effects.
To our knowledge this mechanism of generating FCNCs mediated by {the} $Z$ in
$Z^\prime$ models has not been considered in the literature. Usually the
FCNCs in $Z^\prime$ models are generated through $Z-Z^\prime$ mixing in the
process of the spontaneous breakdown of the electroweak symmetry \cite{Langacker:2008yv}. It is then of interest to investigate whether this pure RG effect is important.

Our paper is organized as follows. In Section~\ref{sec:2} we recall the strategy  of \cite{Buras:2015jaq} where the correlations between
$\epe$, $\Delta M_K$, $\kpn$ and $\klpn$ have been analyzed in the framework of
$Z^\prime$ models taking into account the constraints from $K_{L}\to \mu^+\mu^-$
and $\varepsilon_K$. We refrain, with the exception of $\epe$, from listing the formulae for
observables entering our analysis as they can be found in \cite{Buras:2015jaq}
and in {more general papers on $Z^\prime$ models in \cite{Buras:2012jb} and in
  \cite{Buras:2012dp} that deals with 331 models.}  On the other hand
we discuss in some detail the aspects of new dynamics that enrich
the analysis of  \cite{Buras:2015jaq} through the inclusion of the full
machinery of the SMEFT, in particular of the renormalization group effects
from top Yukawa coupling.

In Section~\ref{sec:3} as a preparation for the numerical analysis we discuss
various $Z^\prime$ scenarios and the related RG evolution patterns in the SMEFT.

In Section~\ref{sec:3a}
we present a detailed numerical analysis of
all observables listed above, including also $K_{S}\to \mu^+\mu^-$ and  $K_L\to\pi^0\ell^+\ell^-$, in various $Z^\prime$ scenarios.

In Section~\ref{sec:4} we analyze the
generation of FCNCs mediated by the SM $Z$-boson.
 In Section~\ref{sec:5} we list
 the main results of our paper and present a brief outlook for the coming years.
Some additional information is contained in an appendix.
  \section{Basic Formalism}\label{sec:2}
\subsection{Strategy}
In our paper, as in  \cite{Buras:2015jaq}, {an important} role will be played by $\epe$ and $\varepsilon_K$ for which in the presence of NP contributions, {to be called BSM in what follows}, we have
\be\label{GENERAL}
\frac{\varepsilon'}{\varepsilon}=\left(\frac{\varepsilon'}{\varepsilon}\right)^{\rm SM}+\left(\frac{\varepsilon'}{\varepsilon}\right)^{\rm BSM}\,, \qquad \varepsilon\equiv\varepsilon_K=
e^{i\phi_\eps}\,
\left[\varepsilon_K^{\rm SM}+\varepsilon^{\rm BSM}_K\right] \,.
\ee
In view of uncertainties present still in the SM estimates of $\epe$, and to a lesser
extent in $\varepsilon_K$, we will fully concentrate on BSM contributions. Therefore in order to identify the pattern of BSM contributions to flavour observables
implied by allowed BSM contributions to $\epe$ in a transparent manner, we will proceed
in a given $Z^\prime$ scenario as follows  \cite{Buras:2015jaq}:

{\bf Step 1:} We assume that BSM provides a  shift in $\epe$:
\be\label{deltaeps}
\left(\frac{\varepsilon'}{\varepsilon}\right)^{\rm BSM}= \kepe\cdot 10^{-3}, \qquad   -1.0 \le \kepe \le 1.0,
\ee
with the range for $\kepe$ indicating conservatively the room left for BSM contributions.
This range is dictated by the recent analyses in \cite{Buras:2020pjp,Aebischer:2020jto}
which implies the result quoted in (\ref{ABB}).
Specifically, we will consider three ranges for $\kepe$
\be \label{eq:epsp_cases}
(\text{A})\quad  0.5 \le \kepe \le 1.0, \qquad (\text{B})\quad -0.5 \le \kepe \le 0.5, \qquad (\text{C})\quad
-1.0 \le \kepe \le -0.5.
\ee
Only range A has been considered in \cite{Buras:2015jaq} so that the
study of ranges B and C is new with interesting consequences.

This step
will determine for given flavour conserving $Z^\prime \bar q q$ couplings the imaginary
parts of flavour-violating $Z^\prime$ couplings to quarks as functions of $\kepe$. But as we will see below in order to explain the  $\Delta M_K$ anomaly, {which requires} significant imaginary couplings, and
  simultaneously obtain $\epe$ consistent with the ranges above the
flavour conserving $Z^\prime \bar q q$ couplings must be $\ord(10^{-2})$.

 We stress even stronger the usefulness of $\kappa_{\varepsilon^\prime}$ in the 2020s than it could be  anticipated in \cite{Buras:2015jaq}.
The result in (\ref{ABB}) governed by the hadronic matrix elements from {the}
  RBC-UKQCD collaboration  has a very large error and we expect that it will still   take some time before this error will be decreased down
   to $10-15\%$.
In addition we need a second lattice group to confirm the 2020 RBC-UKQCD  value and it is not evident that this will happen in this decade.

{\bf Step 2:} In order to determine the relevant real parts of the couplings
involved, in the presence of the imaginary part determined from $\epe$, we will assume that  BSM can also affect the parameter  $\varepsilon_K$.  We will describe
this effect by the parameter $\keps$  so that now in addition
to (\ref{deltaeps}) we will allow for a BSM shift in $\varepsilon_K$ in
the range
\be \label{DES}
(\varepsilon_K)^{\rm BSM}= \keps\cdot 10^{-3},\qquad -0.2\le \keps \le 0.2 \,.
\ee
This is consistent with present analyses in \cite{Bona:2006ah,Charles:2015gya,Brod:2019rzc}. But it should be stressed that
this depends on whether inclusive or exclusive determinations of $\vub$ and
$\vcb$ are used and with the inclusive ones the SM value of $\varepsilon_K$ agrees well with the data. We will also investigate how our results
  change when a larger NP contribution to $\varepsilon_K$ corresponding
  to $-0.5\le \keps \le 0.5$ is admitted.

{\bf Step 3:}  As far as $\Delta M_K$ is concerned, we will consider
  dominantly NP parameters which provide the suppression of the SM value
  in accordance with the LQCD result in (\ref{NPDMK}). In particular this will
  require the imaginary $Z^\prime$ couplings to be significantly larger than the
real ones.

{\bf Step 4:} In view of the uncertainty in $\kepe$ we set several
 parameters  to their central values. In particular
for the SM contributions to rare decays we set the CKM factors and the CKM phase $\delta$ to

\be\label{CKMfixed}
 {\rm Re}\lambda_t=-3.4\cdot 10^{-4}, \qquad {\rm Im}\lambda_t=1.48\cdot 10^{-4}\,, \qquad \delta= 1.27\,,
\ee
which are close to the central values  of present  estimates obtained by the UTfit \cite{Bona:2006ah} and CKMfitter \cite{Charles:2015gya} collaborations.
For this choice of CKM parameters the central value of the resulting { $|\varepsilon_K^{\rm SM}|$ is $2.32\cdot 10^{-3}$. With the experimental value of $\varepsilon_K$ in
 Table~\ref{tab:num1} this implies $\keps=-0.09$ }.  But we will still vary
$\keps$ while keeping the values in (\ref{CKMfixed}), as BSM contributions in our scenarios
do not depend on them but are sensitive functions of $\keps$.

 {\bf Step 5:} Having fixed the flavour violating couplings of {the} $Z^\prime$
in this manner, we will be able to calculate BSM contributions to the branching ratios for $\kpn$, $\klpn$,
$K_{L,S}\to\mu^+\mu^-$ and $K_L\to\pi^0\ell^+\ell^-$
 and to $\Delta M_K$ in terms of $\kepe$ and $\keps$. This
will allow us to study directly the impact of possible NP contributions to $\epe$ and $\Delta M_K$ in $Z^\prime$
scenarios on $\kpn$ {and} $\klpn$ and the remaining rare Kaon decays. In  Table~\ref{tab:IMRE} we indicate the dependence of a given
observable on the {\it real} and/or {\it imaginary} $Z^\prime$ or { later}  $Z$ flavour
violating coupling to quarks.
 In our strategy imaginary parts depend only on $\kepe$ and the choice of flavour
conserving $Z^\prime \bar q q$ couplings, while the real parts {depend}
on both $\kepe$ and $\keps$. The pattern of flavour violation  depends in a given BSM scenario on the
relative size of {the} real and imaginary parts of {the} couplings {as we will see} explicitly later on.

\begin{table}[H]
\begin{center}
\begin{tabular}{|c|c|c|}
\hline
  &  ${\rm Im}\Delta$  & ${\rm Re}\Delta$ \\
  \hline
$\epe$ & $*$ & \\
\hline
$\varepsilon_K$ &  $*$ & $*$ \\
\hline
$\Delta M_K$ & $*$ & $*$ \\
\hline
$\klpn$ & $*$ & \\
\hline
$\kpn$ & $*$ & $*$ \\
\hline
$K_L\to\mu^+\mu^-$ & & $*$\\
\hline
$K_S\to\mu^+\mu^-$ & $*$ & \\
\hline
$K_L\to\pi^0\ell^+\ell^-$ & $*$ &  \\
\hline
\end{tabular}
\end{center}
\captionsetup{width=0.9\textwidth}
\caption{The dependence of various observables on the  imaginary and/or real
parts of $Z^\prime$ and $Z$ flavour-violating couplings.
\label{tab:IMRE}}
\end{table}

In the context of our presentation we will see that in most of our $Z^\prime$ scenarios $\varepsilon_K$ and not $K_L\to \mu^+\mu^-$ is the most important observable
for the determination of the real parts of the new couplings after the $\epe$
constraint has been imposed. This can be traced back to Yukawa RG effects.
 Additional constraint will come from $\Delta M_K$.

\subsection{SMEFT at work}\label{sec:SMEFT}
The interaction Lagrangian of a $Z'=(1,1)_0$ field and the SM fermions reads:
\begin{align}\label{eq:Zplag}
  \mathcal{L}_{Z'}=& -g_q^{ij} (\bar q^i \gamma^\mu q^j)Z'_{\mu}-g_u^{ij} (\bar u^i \gamma^\mu u^j)Z'_{\mu}-g_d^{ij} (\bar d^i \gamma^\mu d^j)Z'_{\mu}  \\\notag
  &-g_\ell^{ij} (\bar \ell^i \gamma^\mu \ell^j)Z'_{\mu}-g_e^{ij} (\bar e^i \gamma^\mu e^j)Z'_{\mu}\,.
\end{align}
Here $q^i$ and $\ell^i$ denote left-handed $SU(2)_L$ doublets and $u^i$, $d^i$ and $e^i$ are right-handed singlets.

This $Z^\prime$ theory will then be matched at the scale $M_{Z^\prime}$ onto the SMEFT, generating the operators listed in Table~\ref{tab:SMEFT-4Fops}.
In
 the Warsaw basis \cite{Grzadkowski:2010es} the tree-level matching \cite{deBlas:2017xtg} with the couplings in (\ref{eq:Zplag}) is given for purely {\em left-handed} vector operators by:
\begin{align}\label{eq:LLmatch}
 \wc[]{\ell\ell}{ijkl}& = -\frac{g_\ell^{ij}g_\ell^{kl}}{2M_{Z'}^2}\,,&
 \wc[(1)]{qq}{ijkl}=& -\frac{g_q^{ij}g_q^{kl}}{2M_{Z'}^2}\,, \\
  \wc[(1)]{\ell q}{ijkl}& = -\frac{g_\ell^{ij}g_q^{kl}}{M_{Z'}^2}\,.
\end{align}

 For purely {\em right-handed}  vector operators one finds:
\begin{align}
\wc[]{ee}{ijkl}=& -\frac{g_e^{ij}g_e^{kl}}{2M_{Z'}^2}\,, &
\wc[]{dd}{ijkl}=& -\frac{g_d^{ij}g_d^{kl}}{2M_{Z'}^2}\,, \\
\wc[]{uu}{ijkl}=& -\frac{g_u^{ij}g_u^{kl}}{2M_{Z'}^2}\,, &
\wc[]{ed}{ijkl}=& -\frac{g_e^{ij}g_d^{kl}}{M_{Z'}^2}\,, \\
\wc[]{eu}{ijkl}=& -\frac{g_e^{ij}g_u^{kl}}{M_{Z'}^2}\,, &
\wc[(1)]{ud}{ijkl}=& -\frac{g_u^{ij}g_d^{kl}}{M_{Z'}^2}\,.
\end{align}

Finally for {\em left-right} vector operators the matching reads:
\begin{align}
\wc[]{\ell e}{ijkl}=& -\frac{g_\ell^{ij}g_e^{kl}}{M_{Z'}^2}\,, &
\wc[]{\ell d}{ijkl}=& -\frac{g_\ell^{ij}g_d^{kl}}{M_{Z'}^2}\,, \\
\wc[]{\ell u}{ijkl}=& -\frac{g_\ell^{ij}g_u^{kl}}{M_{Z'}^2}\,, &
\wc[]{q e}{ijkl}=& -\frac{g_q^{ij}g_e^{kl}}{M_{Z'}^2}\,, \\ \label{eq:LRmatch}
\wc[(1)]{q u}{ijkl}=& -\frac{g_q^{ij}g_u^{kl}}{M_{Z'}^2}\,, &
\wc[(1)]{q d}{ijkl}=& -\frac{g_q^{ij}g_d^{kl}}{M_{Z'}^2}\,.
\end{align}

Different bases\footnote{In the following we adopt the basis conventions defined in \texttt{WCxf} \cite{Aebischer:2017ugx}.} for the SMEFT Wilson coefficients (corresponding to different models) can be used to perform the numerical analysis. A particular choice of basis is the {\em down-basis} \footnote{The down-basis was first discussed in \cite{Aebischer:2015fzz}.}, in which the down-type Yukawas are diagonal and the $q^i$ fields are given above the EW scale by

{
\begin{equation}
	 q^i= \quad
\begin{pmatrix}
	V^\dagger_{ij} u_L^j  \\
d_L^i
\end{pmatrix}
\,, \qquad (\text{down-basis})
\end{equation}
where $V$ denotes the CKM matrix.
Another popular basis choice is the {\em up-basis} with diagonal up-type Yukawas and
\begin{equation}
         q^i= \quad
\begin{pmatrix}
u_L^i  \\
	V_{ij} d_L^j
\end{pmatrix}
\,.\qquad (\text{up-basis})
\end{equation}
 Changing between these two bases is achieved by rotating the corresponding parameters by CKM factors. For instance, to express the up-basis $g_q^{ij}$ couplings in terms of the down-basis ones,
the following rotation needs to be performed:
\begin{equation}
	g_q^{ij} \to (V g_q V^\dagger)^{ij}.
\end{equation}
}
Since we are interested in FCNCs in the down-sector, it is more convenient to work in the down-basis, which we will adopt in the following.
In a next step the SMEFT Wilson coefficients are evolved from the matching scale $\Lambda$ down to the EW scale $\mu_{\rm EW}$. In order to perform this RG evolution the SM parameters are first run up to the high scale $\Lambda$, such that all input parameters (Wilson coefficients and SM parameters) are evolved from the same scale down to $\mu_{\rm EW}$. The procedure to obtain the SM parameters at the high scale is discussed in the next subsection.

\begin{table}[H]
\centering
\renewcommand{\arraystretch}{1.5}
\begin{tabular}{cc|cc}
\toprule
\multicolumn{2}{c|}{$(\bar{L}L)(\bar{L}L)$}&
\multicolumn{2}{c}{$(\bar{R}R)(\bar{R}R)$}
\\
\hline
  $\Op[]{\ell\ell}$     & $(\bar \ell_i \gamma_\mu \ell_j)(\bar \ell_k \gamma^\mu \ell_l)$
& $\Op[]{ee}$          & $(\bar e_i \gamma_\mu e_j)(\bar e_k \gamma^\mu e_l)$
\\
  $\Op[(1)]{qq}$     & $(\bar q_i \gamma_\mu q_j)(\bar q_k \gamma^\mu q_l)$
& $\Op{uu}$          & $(\bar u_i \gamma_\mu u_j)(\bar u_k \gamma^\mu u_l)$
\\
  $\Op[(1)]{\ell q}$ & $(\bar \ell_i \gamma_\mu \ell_j)(\bar q_k \gamma^\mu q_l)$
& $\Op{dd}$          & $(\bar d_i \gamma_\mu d_j)(\bar d_k \gamma^\mu d_l)$
\\
\cline{1-2}
  \multicolumn{2}{c|}{$(\bar{L}L)(\bar{R}R)$}
& $\Op{ed}$          & $(\bar e_i \gamma_\mu e_j)(\bar d_k \gamma^\mu d_l)$
\\
\cline{1-2}
  $\Op{\ell e}$      & $(\bar \ell_i \gamma_\mu \ell_j)(\bar e_k \gamma^\mu e_l)$
& $\Op{eu}$          & $(\bar e_i \gamma_\mu e_j)(\bar u_k \gamma^\mu u_l)$
\\

  $\Op{\ell u}$      & $(\bar \ell_i \gamma_\mu \ell_j)(\bar u_k \gamma^\mu u_l)$
& $\Op[(1)]{ud}$     & $(\bar u_i \gamma_\mu u_j)(\bar d_k \gamma^\mu d_l)$
\\
  $\Op{\ell d}$      & $(\bar \ell_i \gamma_\mu \ell_j)(\bar d_k \gamma^\mu d_l)$
&
\\
  $\Op{qe}$          & $(\bar q_i \gamma_\mu q_j)(\bar e_k \gamma^\mu e_l)$
&
\\
  $\Op[(1)]{qu}$     & $(\bar q_i \gamma_\mu q_j)(\bar u_k \gamma^\mu u_l)$
&
\\
  $\Op[(1)]{qd}$     & $(\bar q_i \gamma_\mu q_j)(\bar d_k \gamma^\mu d_l)$
&
\\
\bottomrule
\end{tabular}
\renewcommand{\arraystretch}{1.0}
  \caption{List of the dimension-six four-fermion ($\psi^4$) operators in SMEFT
  that are generated in a $Z'$ model at tree-level. Flavour
  indices on the quark and lepton fields are $ijkl$.}
  \label{tab:SMEFT-4Fops}
\end{table}

\subsection{Treatment of SM parameters}\label{sec:basisrot}
In order to solve the RGEs, assuming experimental values of the SM parameters at the EW scale,
we evolve them to the input scale $\Lambda$.  For this purpose we employ {an} iterative procedure, which was used in \cite{Aebischer:2018bkb}. This procedure for solving the RGEs
  incorporates the correct values of the CKM parameters and the quark and
  lepton masses at the electroweak scale. Since in the present paper we are interested in exploring the role of {Yukawa RGE effects, let us} describe the iterative procedure to determine the Yukawa couplings at the input scale $\Lambda$:
\begin{itemize}
\item
\noindent
{We start with} the Yukawa matrices in the down-basis at the EW scale:
\begin{equation}
	Y_d = \frac{\sqrt 2}{v}M_d + \frac{\Wc{dH}{} v^2}{2}, \,\ \,\
	Y_u = \frac{\sqrt 2}{v}M_u + \frac{\Wc{uH} v^2}{2}, \,\ \,\
	Y_e = \frac{\sqrt 2}{v}M_e + \frac{\Wc{eH} v^2}{2}\,,
	\label{eq:yukawas}
\end{equation}
with the mass matrices given by
 \begin{equation}
	 \label{eq:smparaew}
M_d =
	 \begin{pmatrix} m_d & 0 & 0\\
	                 0 & m_s & 0 \\
	                 0 & 0 & m_b
	 \end{pmatrix}, \,\ \,\
M_u =
      V^\dagger   \begin{pmatrix} m_u & 0 & 0\\
                         0 & m_c & 0 \\
                         0 & 0 & m_t
         \end{pmatrix}, \,\ \,\
 M_e =
        \begin{pmatrix} m_e & 0 & 0\\
                         0 & m_\mu & 0 \\
                         0 & 0 & m_\tau
         \end{pmatrix}\,.
 \end{equation}
 Here the values of the quark masses can be found in
 Table 1 of \cite{Aebischer:2018bkb}.

\item
In the first step the Yukawa matrices are evolved up to the input scale $\Lambda$ while assuming constant Wilson coefficients (equal to their input values, $C_i(\Lambda)$). As the  chosen basis {is} not stable under RG running, a rotation of the fermion fields is performed to get back to the down-basis.
\begin{equation}\label{eq:rot}
        \psi_f^\prime = U_f \psi_f, \,\ \,\ f = q,u,d,\ell, e,
\end{equation}
taking
\begin{equation}
        U_q = U_{d_L}, \,\ U_d = U_{d_R}, \,\ U_u = U_{u_R}, \,\ U_\ell = U_{e_L},\,\ U_e = U_{e_R} \,.
\end{equation}
Here the unprimed fields are in the down-basis, whereas the primed fields are in some random basis generated by the running of Yukawas just performed.
The rotation matrices at the input scale transform the primed mass matrices back to the down-basis
\begin{eqnarray}
	 M_d(\Lambda) &= & U_{d_L}^\dagger M_d^\prime (\Lambda) U_{d_R}, \\
	 M_u(\Lambda) &= & U_{d_L}^\dagger M_u^\prime (\Lambda) U_{u_R}, \\
	 M_e(\Lambda) &= & U_{e_L}^\dagger M_e^\prime (\Lambda) U_{e_R}\,,
	 \label{eq:rotation}
\end{eqnarray}
obtaining the diagonal matrices $ M_d(\Lambda)$ and $M_e(\Lambda)$ and the non-diagonal matrix $ M_u(\Lambda)$ given in \eqref{eq:smparaew}. The primed matrices at the input scale are given by
\begin{eqnarray}
	M_d^\prime(\Lambda) &=& \frac{v}{\sqrt 2 } \left [ Y_d(\Lambda) -\frac{\Wc{dH }(\Lambda) v^2}{2} \right ], \\
	M_u^\prime(\Lambda) &=& \frac{v}{\sqrt 2 }  \left [Y_u(\Lambda) -\frac{\Wc{uH }(\Lambda) v^2}{2} \right ] \,,\\
	M_e^\prime(\Lambda) &=& \frac{v}{\sqrt 2 }  \left [Y_e(\Lambda) -\frac{\Wc{eH }(\Lambda) v^2}{2} \right ] \,.
\label{eq:}
\end{eqnarray}
\item
In the second step the Wilson coefficients are evolved down to {the} EW
scale in the leading log (LL) approximation, using the Yukawa matrices from the previous step.
\item
Finally, the obtained Yukawas {are evolved up }to the input scale $\Lambda$ using the constant Wilson coefficients {obtained from the LL running}.
\end{itemize}

This iterative procedure allows to find the Yukawa matrices (and other SM parameters) at the high scale. The RGEs can then be solved with all parameters having their initial conditions at the same scale $\Lambda$. We reemphasize that the form of the Yukawa matrices is not stable under RGEs and therefore a back-rotation \cite{Aebischer:2020lsx} is required to go back to the down basis at the EW scale. A crucial consequence of this is that one also needs to
\emph{back-rotate} \cite{Aebischer:2018bkb,Coy:2019rfr,Matsuzaki:2017bpp,Assad_2018,DiLuzio:2017vat,Bordone_2018,DiLuzio:2018zxy} the Wilson coefficients according to Table 4 of \cite{Dedes:2017zog}. We will return to one of these consequences in Section~\ref{rerotation}.

\boldmath
\section{$Z^\prime$ Contributions: Setup}\label{sec:3}
\unboldmath

\subsection{Scenarios}

For the numerical analysis we follow closely the reasoning in \cite{Buras:2015jaq}.
As we are interested in Kaon decays,  we will assume different scenarios for the
flavour transition $d\rightarrow s$, to be referred to as LHS and RHS in the
following. {In these scenarios we allow for a flavour-violating coupling in the
left-handed (LHS) or right-handed (RHS) quark sector between the second and
first generation,  respectively. Moreover, we choose the flavour-diagonal
    first generation quark couplings of both chiralities to be non-vanishing
    in both scenarios}.  With this choice of couplings VLL, VRR as well as VLR operators given in (\ref{eq:LLmatch})-(\ref{eq:LRmatch}) are generated in both scenarios. Furthermore, we define the LR scenario first discussed in \cite{Buras:2014sba,Buras:2014zga}, which is equivalent to the LHS or RHS, but without taking into account constraints from $\varepsilon_K$ and $\Delta M_K$. The justification for
  this procedure is {given as follows. In the LR scenario containing LH as well as RH $Z^\prime$ couplings to SM fermions, left-right $\Delta F=2$ operators are generated at tree-level. Their contributions to the mixing amplitudes $M_{12}^{bs},\, M_{12}^{bd}$ and $M_{12}^{sd}$ are RG enhanced. For the $d\to s$ transition there is an additional chiral enhancement. However, by imposing a fine-tuning between the left-left and right-right contributions and the LR contributions, the constraints from $\varepsilon_K$ and $\Delta M_K$ can be alleviated while giving sizable contributions to $K\to\pi\pi$, as has been shown in \cite{Buras:2014sba}. The generalization of this idea to the other meson systems has been done in \cite{Buras:2014zga}.
 We will briefly return to this scenario in Section~ \ref{LRSCENARIO}. For further details we refer to Appendix A of \cite{Aebischer:2019blw}.}

  In the LHS, the flavour change is achieved by the non-zero complex coupling $g_q^{21}$ and in the RHS by a complex-valued $g_d^{21}$. In each scenario we allow for diagonal (real) couplings to first generation quarks. Furthermore, to also accommodate for the decays $K_{L,S}\rightarrow \mu^+\mu^-$ a real non-zero value of $g_\ell^{22}$
is chosen. For $K_L\to\pi^0\ell^+\ell^-$ with $\ell=e,\mu$ we
    also need non-zero  { $g_{\ell}^{11}$ and} $g_\ell^{22}$.  All other couplings are assumed to vanish. Therefore we have at the high scale $\Lambda$ the following three setups:
\begin{align}
\text{LHS}:& \quad g_q^{11,21}\,,\quad g_u^{11}\,,\quad g_d^{11}\,,\quad g_l^{11}\,,\quad g_l^{22}\,, \\
\text{RHS}:& \quad g_q^{11}\,,\quad g_u^{11}\,,\quad g_d^{11,21}\,,\quad g_l^{11}\,,\quad g_l^{22}\,,\\
\text{LR}:& \quad g_q^{11,21}\,,\quad g_u^{11}\,,\quad g_d^{11,21}\,,\quad g_l^{11}\,,\quad g_l^{22}\,.
\end{align}

Such scenarios are in general subject to gauge anomalies, which are assumed to be canceled by additional heavy fields at a higher scale \cite{Langacker:2008yv,Alonso:2018bcg,Smolkovic:2019jow}. $Z'$ models with explicit gauge anomaly cancellation were discussed recently in \cite{Aebischer:2019blw,Altmannshofer:2019xda,DAmico:2017mtc}.

Using the matching relations in Sec.~\ref{sec:SMEFT} leads to the following non-zero {four-fermion} Wilson coefficients in the three different scenarios at the BSM scale:

\begin{align}\label{eq:mathLHS}
\text{LHS}:& \quad \wc[]{\ell\ell}{1111}\,,\wc[]{\ell\ell}{1122}\,,\wc[]{\ell\ell}{2222}\,, \wc[(1)]{\ell q}{1111}\,, \wc[(1)]{\ell q}{1121}\,, \wc[(1)]{\ell q}{2211}\,, \wc[(1)]{\ell q}{2221}\,,\\\notag
&\quad \wc[(1)]{qq}{1111}\,,\wc[(1)]{qq}{1121}\,, \wc[(1)]{qq}{2121}\,, \wc[]{dd}{1111}\,,\wc[]{uu}{1111}\,,\wc[(1)]{ud}{1111}\,,\wc[]{\ell d}{1111}\,,  \\ \notag
&\quad \wc[]{\ell d}{2211}\,,\wc[]{\ell u}{1111}\,, \wc[]{\ell u}{2211}\,, \wc[(1)]{qu}{1111}\,, \wc[(1)]{qu}{2111}\,, \wc[(1)]{qd}{1111}\,, \wc[(1)]{qd}{2111}\,, \\
\text{RHS}:& \quad \text{LHS}\,(g_q^{21}\to g_d^{21})\,,\\
\text{LR}:& \quad \wc[(1)]{qd}{2121}\,,\,\text{LHS}\,,\,\text{RHS}\,.\label{eq:mathLR}
\end{align}

\subsection{RG Running}
The Wilson coefficients obtained in \eqref{eq:mathLHS}-\eqref{eq:mathLR} are then run
down to the EW scale by solving the full set of SMEFT
RGEs \cite{Jenkins:2013zja,Jenkins:2013wua,Alonso:2013hga}. To visualize this
effect different flow charts are shown in Figs.~\ref{fig:run-nonlep}-\ref{fig:run-sl}. We show the
charts of the running of four-fermi operators into operators contributing to
non-leptonic $\Delta S=1$ and  $\Delta S=2$ observables and semi-leptonic
$\Delta S=1$ decays. We present the charts for LHS
and RHS, where LH and RH refer to flavour-violating currents. The structure of
these charts is as follows:
\begin{figure}[htb]
\begin{center}
 \includegraphics[clip, trim=0.5cm 12cm 0.5cm 12cm,width=1.\textwidth]{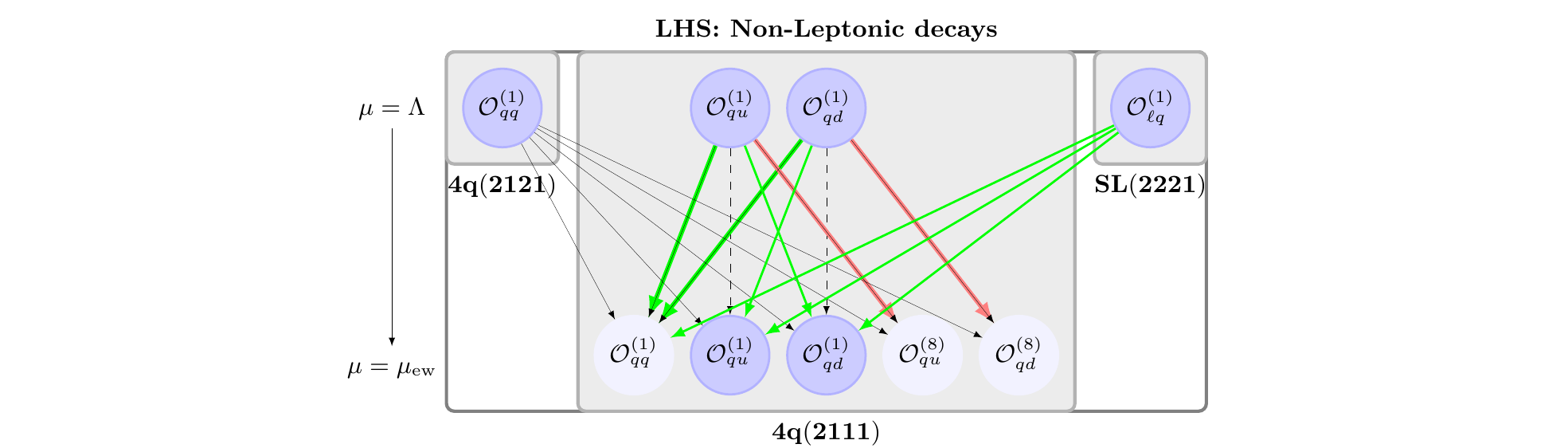}
  \includegraphics[clip, trim=0.5cm 12cm 0.5cm 12cm,width=1.\textwidth]{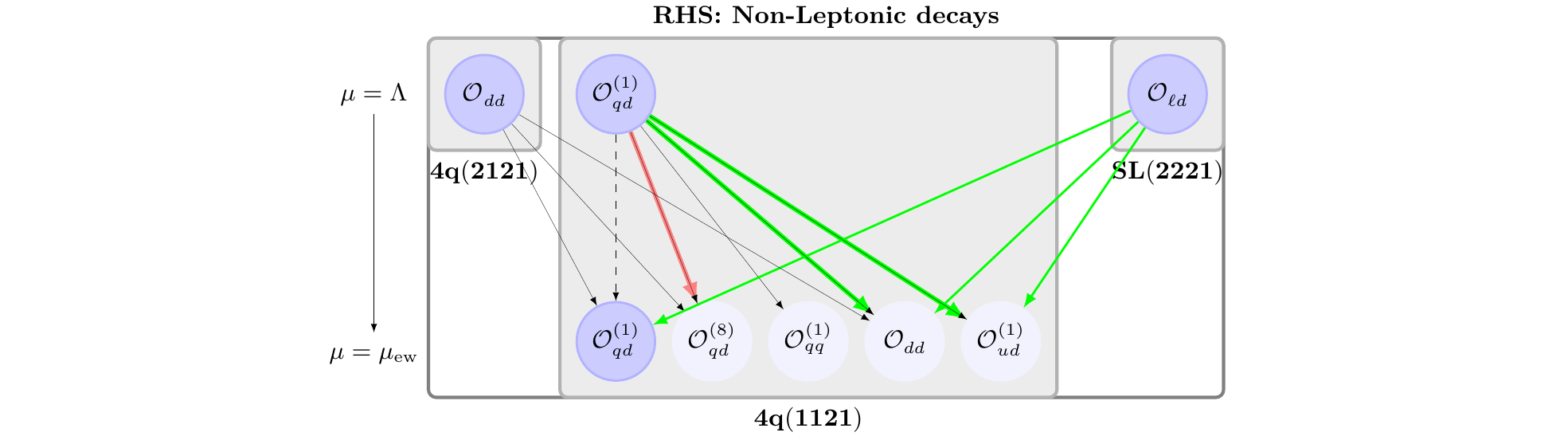}
\captionsetup{width=0.9\textwidth}
\caption{Running of four-fermion operators into operators contributing to $\Delta F=1$ non-leptonic observables. Here the red, green and black lines indicate the operator mixing due to strong, weak and Yukawa couplings respectively. The self-mixing for all couplings is shown by a dashed black line.}
\label{fig:run-nonlep}
\end{center}
\end{figure}
\begin{figure}[htb]
\centering
 \includegraphics[clip, trim=0.5cm 12cm 0.5cm 12cm,width=1.\textwidth]{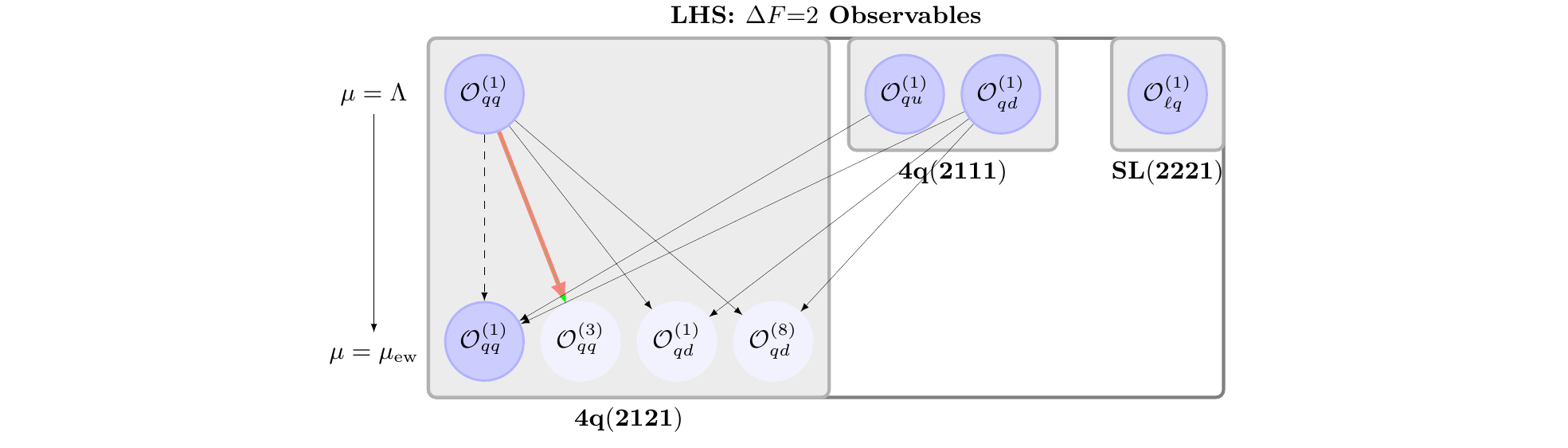}
  \includegraphics[clip, trim=0.5cm 12cm 0.5cm 12cm,width=1.\textwidth]{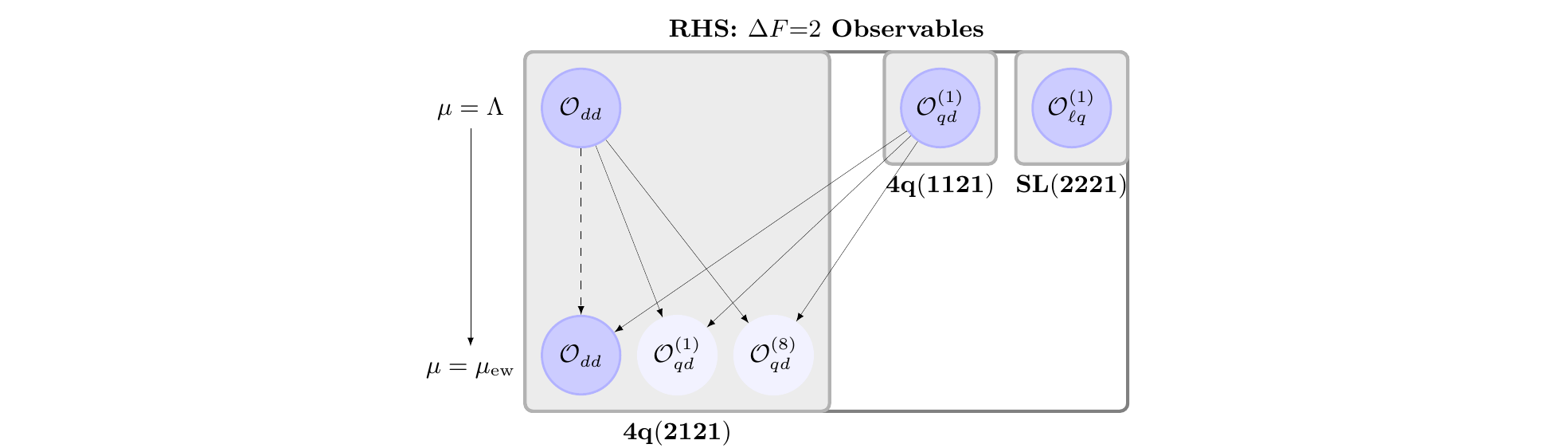}
\captionsetup{width=0.9\textwidth}
\caption{Running of four-fermion operators into operators contributing to $\Delta F=2$ observables. Here the red, green and black lines indicate the operator mixing due to strong, weak and Yukawa couplings respectively. The self-mixing for all couplings is shown by a dashed black line.}
\label{fig:run-ds2}
\end{figure}
\begin{figure}[htb]
 \includegraphics[clip, trim=0.5cm 12cm 0.5cm 12cm,width=1.\textwidth]{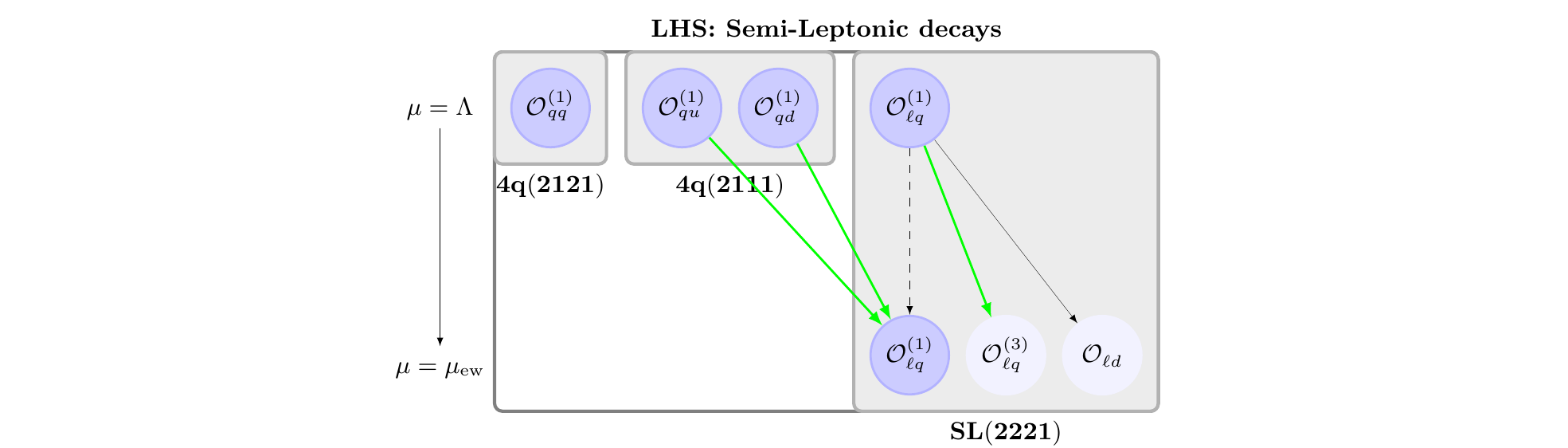}
  \includegraphics[clip, trim=0.5cm 12cm 0.5cm 12cm,width=1.\textwidth]{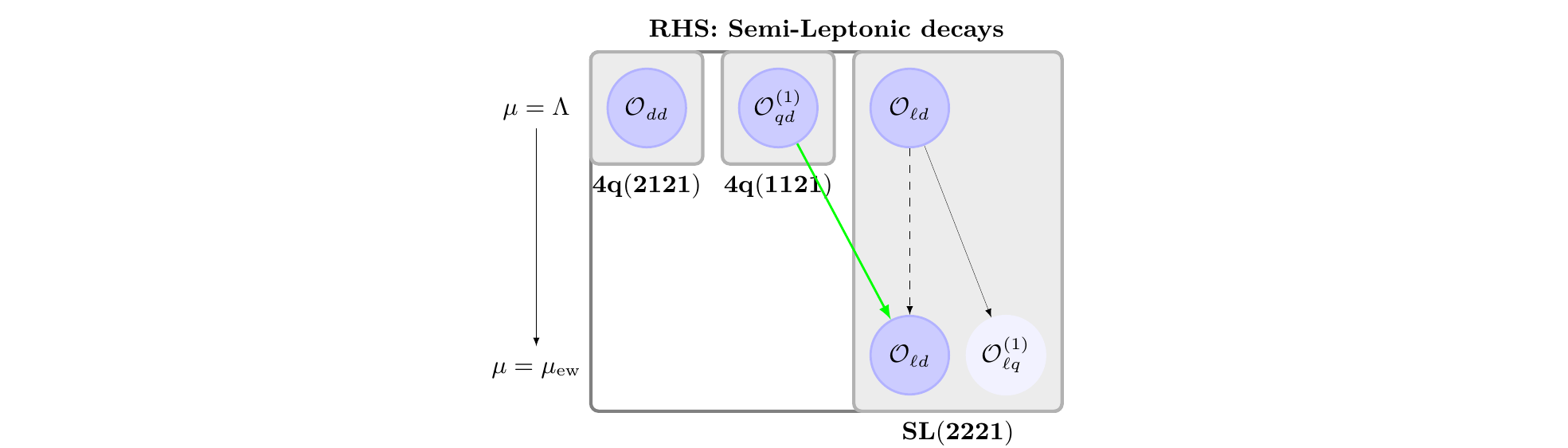}
\captionsetup{width=0.9\textwidth}
\caption{Running of four-fermion operators into operators contributing to $\Delta F=1$ semi-leptonic observables. Here the red, green and black lines indicate the operator mixing due to strong, weak and Yukawa couplings respectively. The self-mixing for all couplings is shown by a dashed black line.}
\label{fig:run-sl}
\end{figure}

\begin{itemize}
\item
  At the BSM scale those operators are listed which on the one hand receive a
non-vanishing matching contribution and on the other hand imply
  through RG evolution contributions at the electroweak scale. The latter can come from the same operators with modified Wilson coefficients and from new operators generated through
  RG evolution. These new operators are placed on a lighter background than the
  original operators.
\item
  As an example consider the first chart in Fig.~\ref{fig:run-nonlep}. The goal is to generate
  at the electroweak scale four-quark operators contributing to non-leptonic
  $\Delta S=1$ processes which is indicated by the indices $(2111)$ . The operators
  \be
  \Op[(1)]{qq}, \qquad \Op[(1)]{qu}, \qquad \Op[(1)]{qd}
  \ee
  present already at the BSM scale contribute also at the electroweak scale but
  whereas  the indices of the Wilson coefficients $\Wc[(1)]{qu}$ and $\Wc[(1)]{qd}$ at the BSM and
  EW scale are the same, the ones of $ \Wc[(1)]{qq}$ change from $(2121)$ to
  $(2111)$.

  In addition the operators $\Op[(8)]{qu}$ and $\Op[(8)]{qd}$ are generated through QCD interactions at the EW scale. Finally the semi-leptonic operator $\Op[(1)]{\ell q}$, present
  already at the BSM scale, while not contributing directly to $\Delta S=1$
  non-leptonic observables, can do it indirectly via Wilson coefficients of non-leptonic operators through electroweak interactions.
\item
  The same logic is used in the remaining charts. But one should note that
in the RHS the flavour-violating indices are on the right-handed currents
  so that e.g. on the top of {the} lower charts {in} Fig.~\ref{fig:run-nonlep}-\ref{fig:run-sl}
the indices are now $(1121)$ instead of $(2111)$.
\item
  The distinction between strong, weak and Yukawa interactions is made with
  the help of colours as described in the figure caption.
\end{itemize}

\subsection{$\epe$}
{Since $\epe$ is one of the key observables in our analysis we discuss here explicitly the impact of the LHS and RHS on this observable.} The relevant SMEFT matching contributions for $\epe$ can be found in \cite{Aebischer:2018csl}. Adopting the same short distance basis as therein, namely
\begin{align}
  \label{eq:DS1-psi4}
  O_{XAB}^q &
  = (\bar s^i \Gamma_X P_A d^i) (\bar q^j \Gamma_X P_B q^j) \,,
&
  \widetilde{O}_{XAB}^q &
  = (\bar s^i \Gamma_X P_A d^j) (\bar q^j \Gamma_X P_B q^i) \,,
\end{align}
with colour indices $i,j$, chiralities $A,B=L,R$, and Dirac structures $X=S,V,T$ with $\Gamma_S=1$,
$\Gamma_V=\gamma^\mu$, $\Gamma_T=\sigma^{\mu\nu}$, one finds at the high scale $\Lambda$:
\begin{align}\label{eq:WETmatchu}
{\rm(LHS)}:&\quad  C_{VLR}^u = \wc[(1)]{qu}{2111} \,, \qquad \qquad C_{VLR}^d = \wc[(1)]{qd}{2111}\,, \\\label{eq:WETmatchd}
{\rm(RHS)}:&\quad  C_{VRL}^u = |V_{ud}|^2\wc[(1)]{qd}{1121} \,, \qquad C_{VRL}^d = \wc[(1)]{qd}{1121}\,,
  \end{align}
where we have neglected small contributions. However, as indicated by the red arrows in Fig.~\ref{fig:run-nonlep}, the Wilson coefficients $\wc[(8)]{qu}{2111}$ and $\wc[(8)]{qd}{2111}$ $(\wc[(8)]{qd}{1121})$ are induced through QCD running down to {the} EW scale in the LHS (RHS). At LL one finds \cite{Celis:2017hod,Alonso:2013hga}:

\begin{equation}
\wc[(8)]{qu}{2111}(\mu_{\rm EW})=-3\frac{\alpha_s}{\pi}\wc[(1)]{qu}{2111}(\Lambda)\ln{\left(\frac{\mu_{\rm EW}}{\Lambda}\right)}\,,
\end{equation}
and similar expressions for $\wc[(8)]{qd}{2111}$ and $\wc[(8)]{qd}{1121}$.  Therefore, taking QCD RGE effects into account the matching at the BSM scale in (\ref{eq:WETmatchu})-(\ref{eq:WETmatchd}) is modified  at the EW scale as
  follows
\begin{align}\label{eq:fullWETmatchLHS1}
{\rm(LHS)}:&\quad  C_{VLR}^u = \wc[(1)]{qu}{2111}-\frac{1}{6}\wc[(8)]{qu}{2111} \,, \qquad C_{VLR}^d = \wc[(1)]{qd}{2111}-\frac{1}{6}\wc[(8)]{qd}{2111}\,, \\
&\quad  \widetilde{C}_{VLR}^u = \frac{1}{2}\wc[(8)]{qu}{2111} \,, \qquad C_{SRL}^d = -\wc[(8)]{qd}{2111}\,, \\
{\rm(RHS)}:&\quad  C_{VRL}^u = |V_{ud}|^2 \big (\wc[(1)]{qd}{1121}-\frac{1}{6}\wc[(8)]{qd}{1121} \big )\,, \quad C_{VRL}^d = \wc[(1)]{qd}{1121}-\frac{1}{6}\wc[(8)]{qd}{1121}\,, \\ \label{eq:fullWETmatchRHS2}
&\quad  \widetilde{C}_{VRL}^u = \frac{1}{2}|V_{ud}|^2\wc[(8)]{qd}{1121} \,, \qquad C_{SLR}^d = -\wc[(8)]{qd}{1121}\,.
  \end{align}

 Employing now the master formula for the BSM contribution {to} $\epe$ one finds \cite{Aebischer:2018rrz,Aebischer:2018csl,Aebischer:2018quc}:
\begin{align}\label{eq:masterWET}
  \left(\frac{\varepsilon'}{\varepsilon}\right)_{\rm BSM} \approx&  -124 \cdot {\rm Im}[ C_{VLR}^u-C_{VRL}^u]+117\cdot  {\rm Im}[C_{VLR}^d-C_{VRL}^d]\ \\\notag
  & -430 \cdot {\rm Im}[ \widetilde{C}_{VLR}^u-\widetilde{C}_{VRL}^u]+204\cdot  {\rm Im}[C_{SLR}^d-C_{SRL}^d] \\\label{eq:masterSMEFT}
  & = {\rm Im}[-124 \wc[(1)]{qu}{2111}-194.3\wc[(8)]{qu}{2111}+117\wc[(1)]{qd}{2111}+184.5\wc[(8)]{qd}{2111} \\\notag
  &+(124 \wc[(1)]{qd}{1121}+194.3\wc[(8)]{qd}{1121})|V_{ud}|^2-117\wc[(1)]{qd}{1121}-184.5\wc[(8)]{qd}{1121}]\,,
\end{align}
where we have used (\ref{eq:fullWETmatchLHS1})-(\ref{eq:fullWETmatchRHS2})  and the Wilson coefficients on the right-hand side of (\ref{eq:masterWET}) and (\ref{eq:masterSMEFT}) are given in units\footnote{ See footnote 7 in \cite{Aebischer:2018csl}.} of ($1/\text{TeV}^2$).
The first and second line in (\ref{eq:masterSMEFT}) correspond to contributions from the LHS and RHS respectively.

\section{$Z^\prime$ Contributions: Numerics}\label{sec:3a}

{In our numerical analysis we investigate the following quantities:}
\begin{eqnarray}\label{eq:kappas}
R_{\Delta M_K}    &=&  \frac{\Delta M_K^{BSM}} {\Delta M_K^{exp}} \,,\quad
R_{\nu\bar\nu}^+  =  \frac{\mathcal{B}(K^+ \to \pi^+ \nu \bar \nu)}{\mathcal{B}(K^+ \to \pi^+ \nu \bar \nu)_{SM}} \,,\quad
R_{\nu\bar\nu}^0 =  \frac{\mathcal{B}(K_L \to \pi^0 \nu \bar \nu)}{\mathcal{B}(K_L \to \pi^0 \nu \bar \nu)_{SM}}\,, \\\notag
R_{\mu^+\mu^-}^S  &=&  \frac{\mathcal{B}(K_S \to \mu^+\mu^-)}{\mathcal{B}(K_S \to \mu^+\mu^-)_{SM}} \,,\quad R_{\pi\ell^+\ell^-}^0 =  \frac{\mathcal{B}(K_L \to \pi^0 \ell^+ \ell^-)}{\mathcal{B}(K_L \to \pi^0 \ell^+ \ell^-)_{SM}}\,.
\end{eqnarray}

\noindent
For the numerical analysis the input parameters in Tables~\ref{tab:num1} and \ref{tab:num2}
are used. The constraint from $\mathcal{B}(K_L\to \mu^+ \mu^-)$ at the 2$\sigma$ level is
taken into account. The SM predictions for $\kpn$ and $\klpn$ are
  given in (\ref{KSM}) and for the remaining decays one finds
\cite{Bobeth:2016llm, Isidori:2003ts, DAmbrosio:2017klp, Mescia:2006jd}:
\newline
\begin{align}\notag
  &\mathcal{B}(K_S\to\mu^+\mu^-)_{SM} = (5.2\pm 1.5)\times 10^{-12}\,,\quad \mathcal{B}(K_L\to\pi^0e^+e^-)_{SM} = 3.54^{+0.98}_{-0.85}(1.56^{+0.62}_{-0.49})\times 10^{-11}\,,\\
  &\mathcal{B}(K_L\to\pi^0\mu^+\mu^-)_{SM} = 1.41^{+0.28}_{-0.26}(0.95^{+0.22}_{-0.21})\times 10^{-11}\,,
\end{align}
where for the $K_L\to\pi^0\ell^+\ell^-$ decays the numbers in parenthesis denote the destructive interference case.
{The experimental status of these decays is given by\cite{AlaviHarati:2003mr, AlaviHarati:2000hs, Aaij:2017tia}}:
\newline
\begin{align}\notag
  &\mathcal{B}(K_S\to\mu^+\mu^-)_{\rm LHCb} < 0.8(1.0) \times 10^{-9}\,,\quad \mathcal{B}(K_L\to\pi^0e^+e^-)_{exp} < 28 \times 10^{-11}\,,\\
  &\mathcal{B}(K_L\to\pi^0\mu^+\mu^-)_{exp} < 38 \times 10^{-11}\,,
\end{align}
Finally, for the LHS and RHS we impose the constraint from $\varepsilon_K$ in the following way:
\begin{eqnarray} \label{eq:kappa_e}
\kappa_{\varepsilon} \in [-0.2,  0.2]\,,
\end{eqnarray}
where $\kappa_{\varepsilon}$ is defined in (\ref{DES}).  But we will investigate what happens for a larger range $\kappa_{\varepsilon} \in [-0.5,  0.5]$.

\begin{table}[tbp]
\centering
{\small
\begin{tabular}{|l|l|l|}
\hline
$G_F = 1.16637(1)\times 10^{-5}\gev^{-2}$\hfill         &  $M_Z = 91.188(2) \gev$\hfill
& $M_W = 80.385(15) \gev$\\
$\sin^2\theta_W = 0.23116(13)$\hfill & $\alpha(M_Z) = 1/127.94$\hfill & $\alpha_s(M_Z)= 0.1184(7) $\\
\hline
$m_e=0.511\mev$\hfill & $m_\mu=105.66\mev$\hfill & $m_\tau=1776.9(1)\mev$\\
$m_u(2\gev)=2.16(11)\mev $ \hfill & $m_c(m_c) = 1.279(13) \gev$ \hfill
& $m_t(m_t) = 163(1)\gev$\\
$m_d(2\gev)=4.68(15)\mev$\hfill & $m_s(2\gev)=93.8(24) \mev$\hfill &
$m_b(m_b)=4.19^{+0.18}_{-0.06}\gev$\\
\hline
$m_{K^\pm}=493.68(2)\mev$\hfill & $m_{K^0}=497.61(1)\mev$\hfill & $\Delta M_K= 0.5292(9)\times 10^{-2} \,\text{ps}^{-1}$
\\
$m_{B_d}=5279.62(15)\mev$\hfill &
$m_{B_s} = 5366.82(22)\mev$\hfill & $|\eps_K|= 2.228(11)\times 10^{-3}$
\\
\hline
\end{tabular}}
\caption{Values of theoretical quantities used for the numerical analysis.}\label{tab:num1}
\end{table}

\begin{table}[H]
\centering
\renewcommand{\arraystretch}{1.3}
\begin{tabular}{|l|l|l|}
\hline
$F_{B_d}$ = $190.5(1.3)\mev$ \hfill &
$F_{B_s}$ = $230.7(1.2)\mev$ \hfill &
$F_K = 156.1(11)\mev$\hfill\\
$\hat B_{B_d} =1.27(10)$ \hfill &  $\hat B_{B_s} =1.33(6)$\hfill
& $\hat B_K= 0.766(10)$ \\
$F_{B_d} \sqrt{\hat B_{B_d}} = 216(15)\mev$\hfill &
$F_{B_s} \sqrt{\hat B_{B_s}} = 266(18)\mev$\hfill &
$\xi = 1.21(2)$\\
$\eta_{cc}=1.87(76)$\hfill & $\eta_{ct}= 0.496(47)$\hfill &
$\eta_{tt}=0.5765(65)$\\
$\eta_B=0.55(1)$\hfill & $\phi_{\varepsilon}=43.51(5)^\circ$\hfill & $\kappa_\varepsilon=0.94(2)$ \\
$|V_{us}|=0.2248(8)$\hfill & $|V_{ub}|=3.73(14)\times10^{-3}$\hfill
&$|V_{cb}|=4.221(78)\times10^{-2}$\\
\hline
\end{tabular}
\caption{{Constants used for the numerical analysis. }}\label{tab:num2}
\end{table}

\subsection{Electroweak Penguin Scenario: Left-Handed}\label{subsec:LHSEWP}
We start with a LHS (i.e. $g_q^{21}\neq 0$), where the effect in $\epe$ is achieved through electroweak
penguin (EWP) operators such as $Q_8$. To generate such operators we choose the quark couplings in the following way:
\begin{equation}
g_q^{21}\neq 0\,, \qquad  g_u^{11}=-2g_d^{11}\,, \qquad g_{\ell}^{11,22} \neq0 \qquad \text{(LH-EWP scenario)}\,.
\end{equation}

\begin{figure}[htb]
\centering
\includegraphics[width=0.4\textwidth]{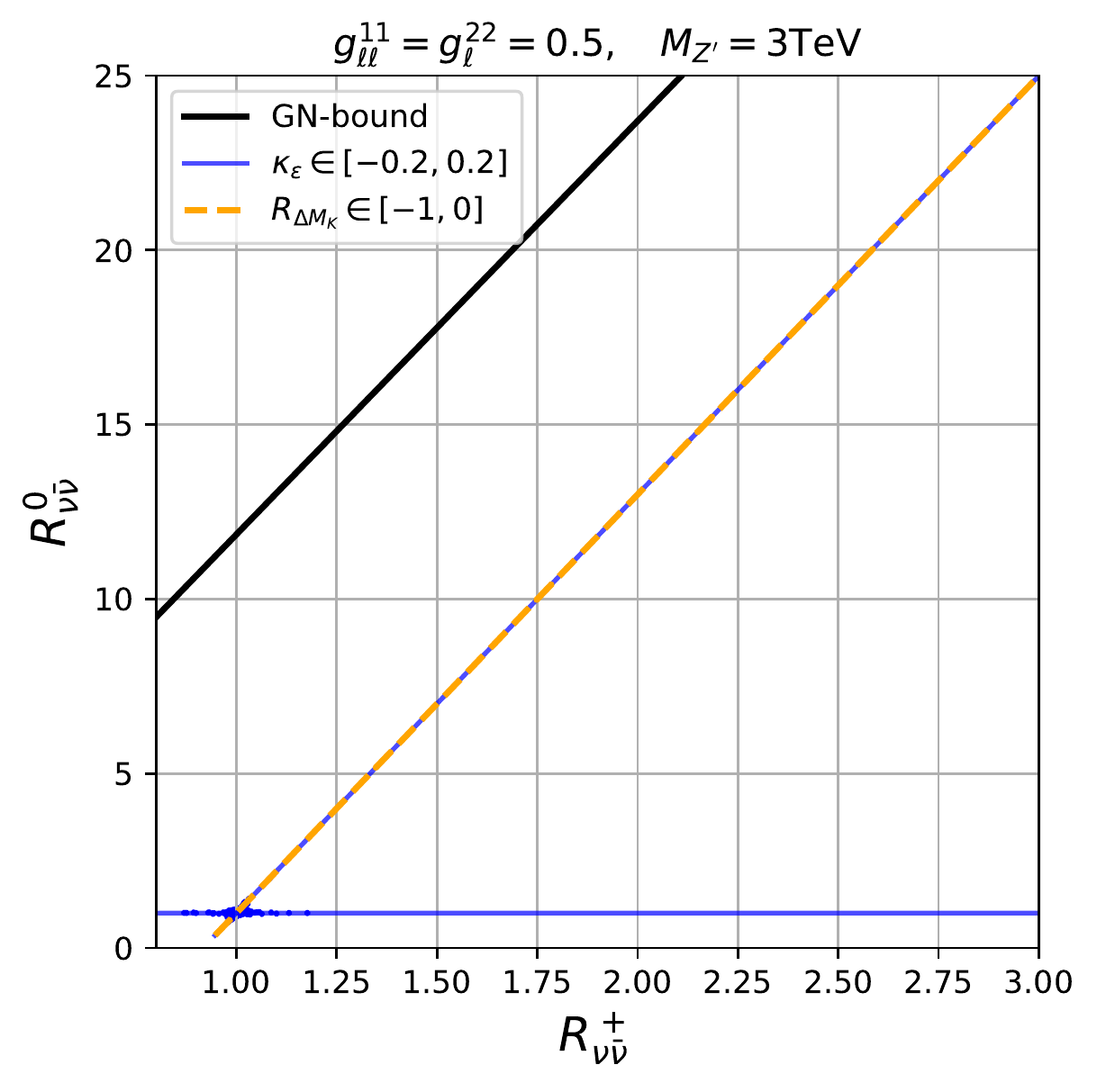}
\includegraphics[width=0.4\textwidth]{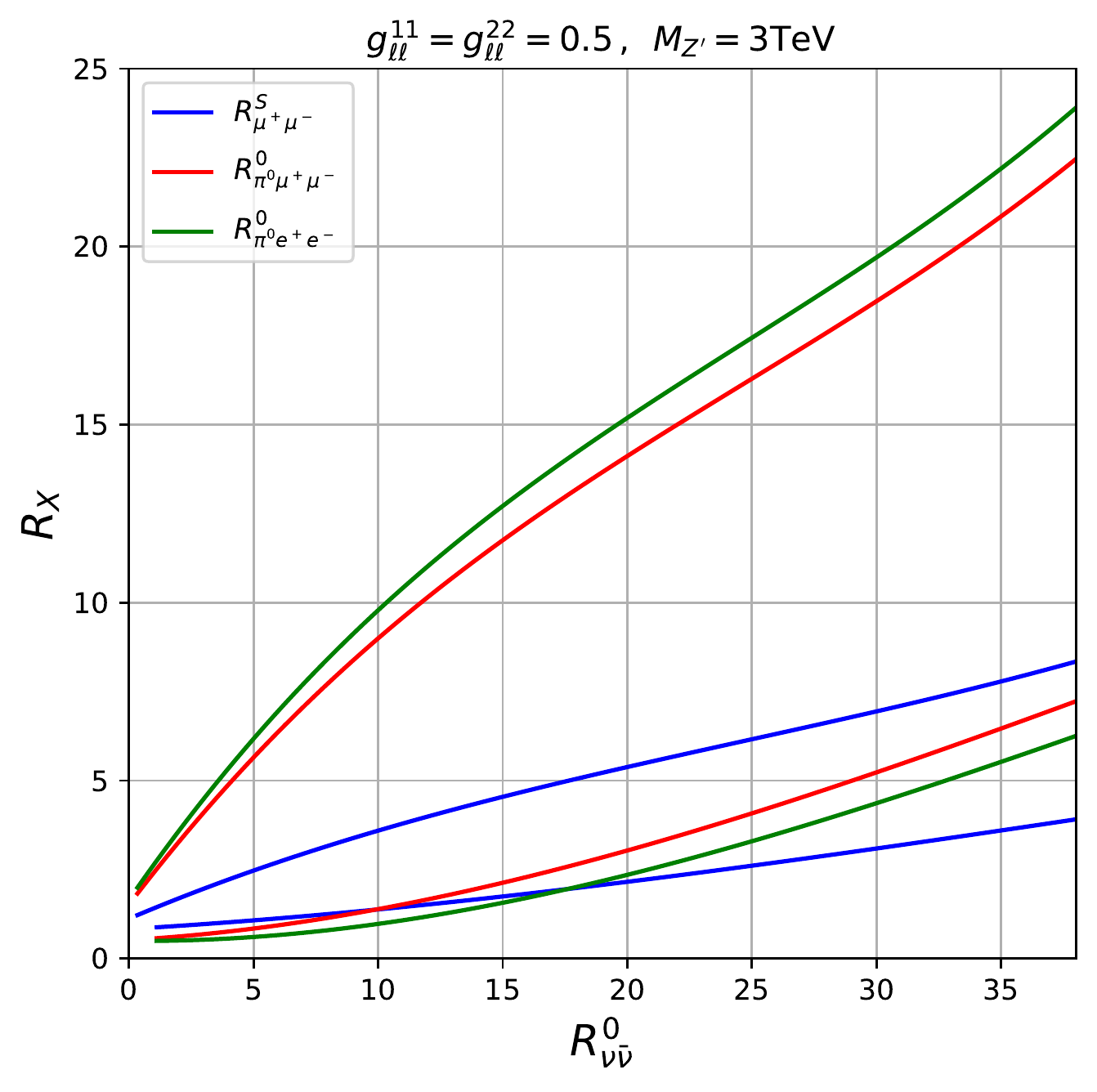}
\captionsetup{width=0.9\textwidth}
\caption{ {LH-EWP scenario for a $Z'$ of $3\tev$. The correlation between the
	ratios for the process $K^+\to\pi^+\nu\bar\nu$, $K_L\to\pi\nu\bar\nu$ defined in \eqref{eq:kappas}}
	is plotted (left). The blue (orange) lines are allowed by $\kappa_{\varepsilon}$
	($\Delta M_K$) constraints {and the black line represents the GN bound}. The correlations between the
	ratio for $K_L\to \pi^0\nu\bar\nu$ and the ones for
	$K \to\pi \ell^+ \ell^-$ and  $K_S \to \mu^+ \mu^-$ are shown (right). }
\label{fig:LHS_kp_k0}
\end{figure}

 In Fig.~\ref{fig:LHS_kp_k0} (left), we plot the correlation between the ratios for the
decays $K^+\to\pi^+\nu\bar\nu$ and $K_L\to\pi\nu\bar\nu$. Here the
horizontal and vertical branches correspond to
purely real and imaginary values respectively of the flavour violating
coupling $g_q^{21}$. Simultaneous presence of both real and
imaginary parts, which correspond to the small area at
the meeting point of {the} two branches, are strongly constrained
by the allowed range of $\kappa_\varepsilon$ (\ref{eq:kappa_e}). {Furthermore},
requiring the suppression of $\Delta M_K$ excludes the horizontal branch, indicating
the dominance of {the} imaginary part over the real {part} of $g_q^{21}$.

\begin{figure}[htb]
\centering
\includegraphics[width=0.4\textwidth]{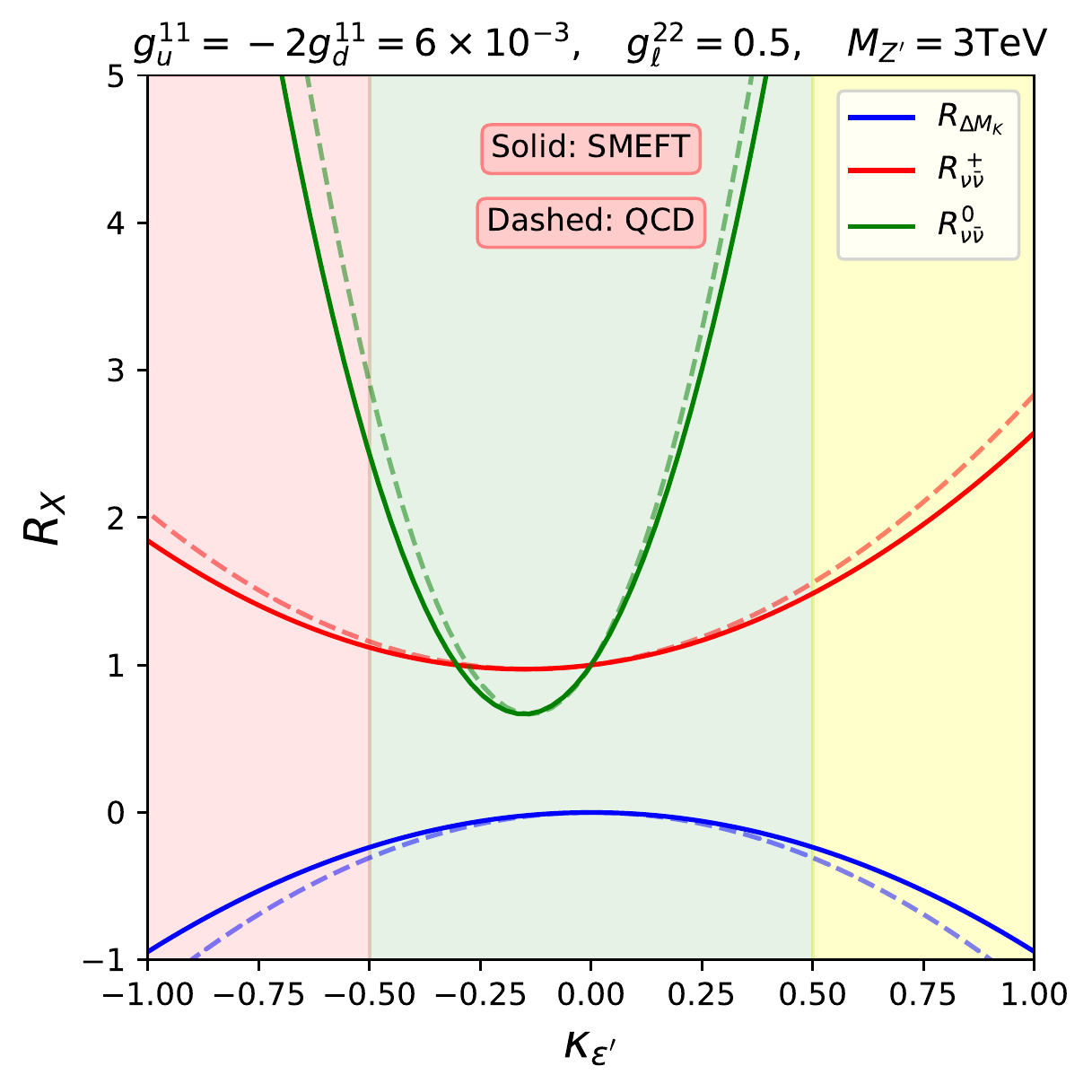}
\includegraphics[width=0.4\textwidth]{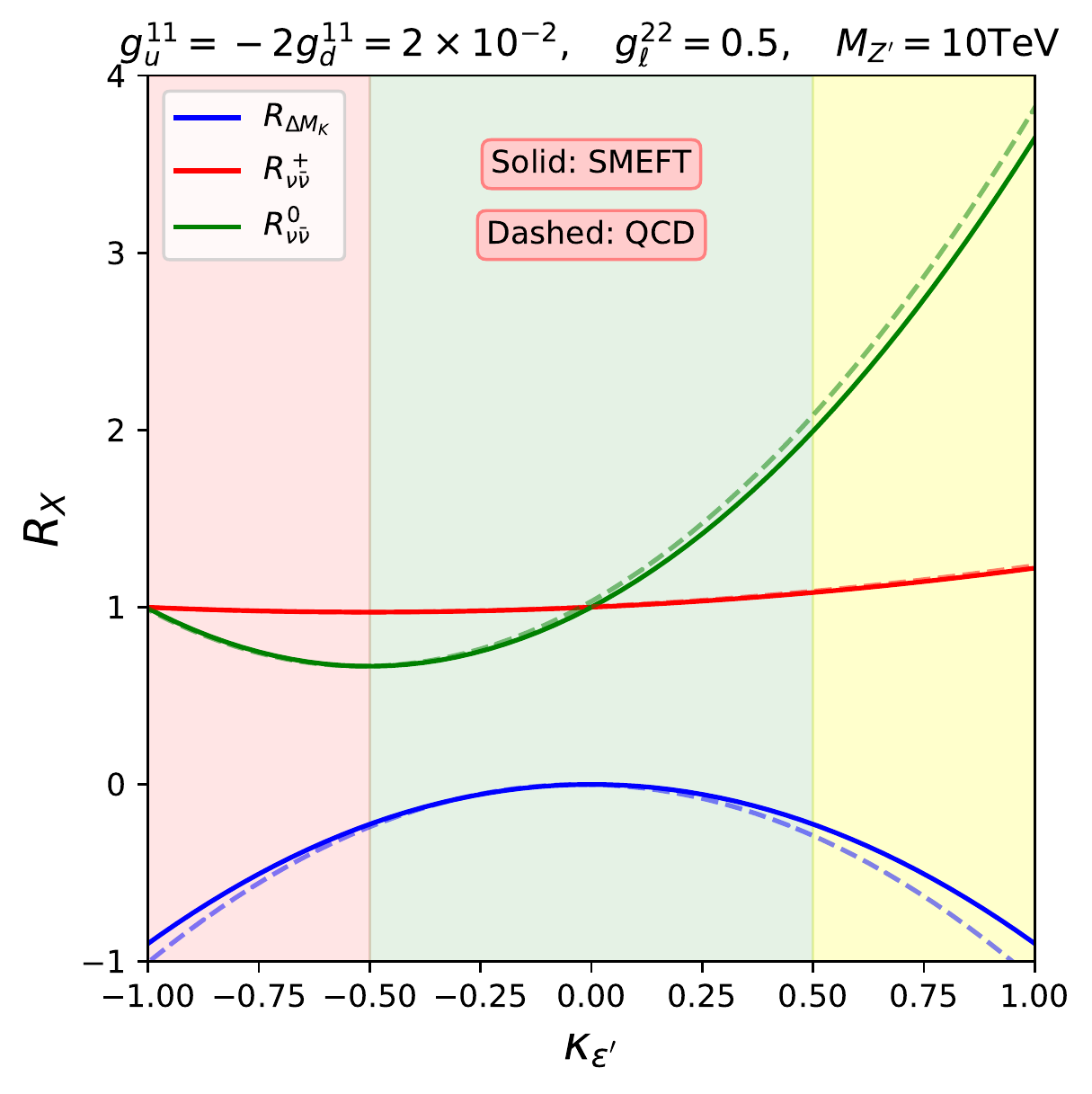}
\captionsetup{width=0.9\textwidth}
\caption{{LH-EWP scenario for a $Z'$ of $3\tev$ (left panel) and $10\tev$ (right panel). The ratios for $\Delta M_K$ and for the process $K^+\to\pi^+\nu\bar\nu$, $K_L\to\pi\nu\bar\nu$ defined in \eqref{eq:kappas}} are plotted against $\kappa_{\varepsilon'}$. The dashed (solid) lines result from QCD (full SMEFT) running above the EW scale. The yellow, green and red shades correspond to
the $\kappa_{\varepsilon^\prime}$ scenarios A, B and C as defined in \eqref{eq:epsp_cases}. }
\label{fig:LHS_kappas1}
\end{figure}
\begin{figure}[H]
\centering
\includegraphics[width=0.4\textwidth]{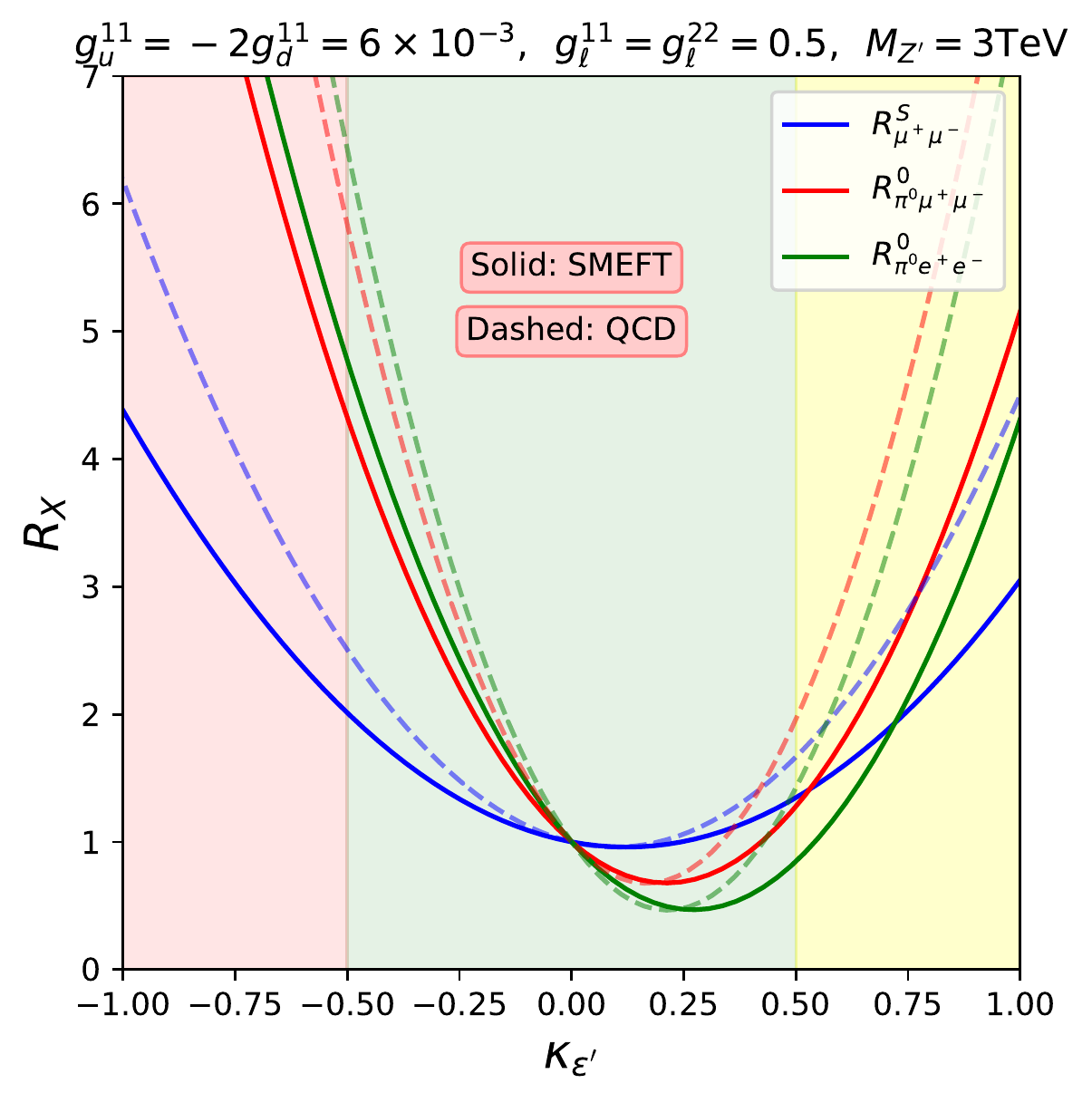}
\includegraphics[width=0.4\textwidth]{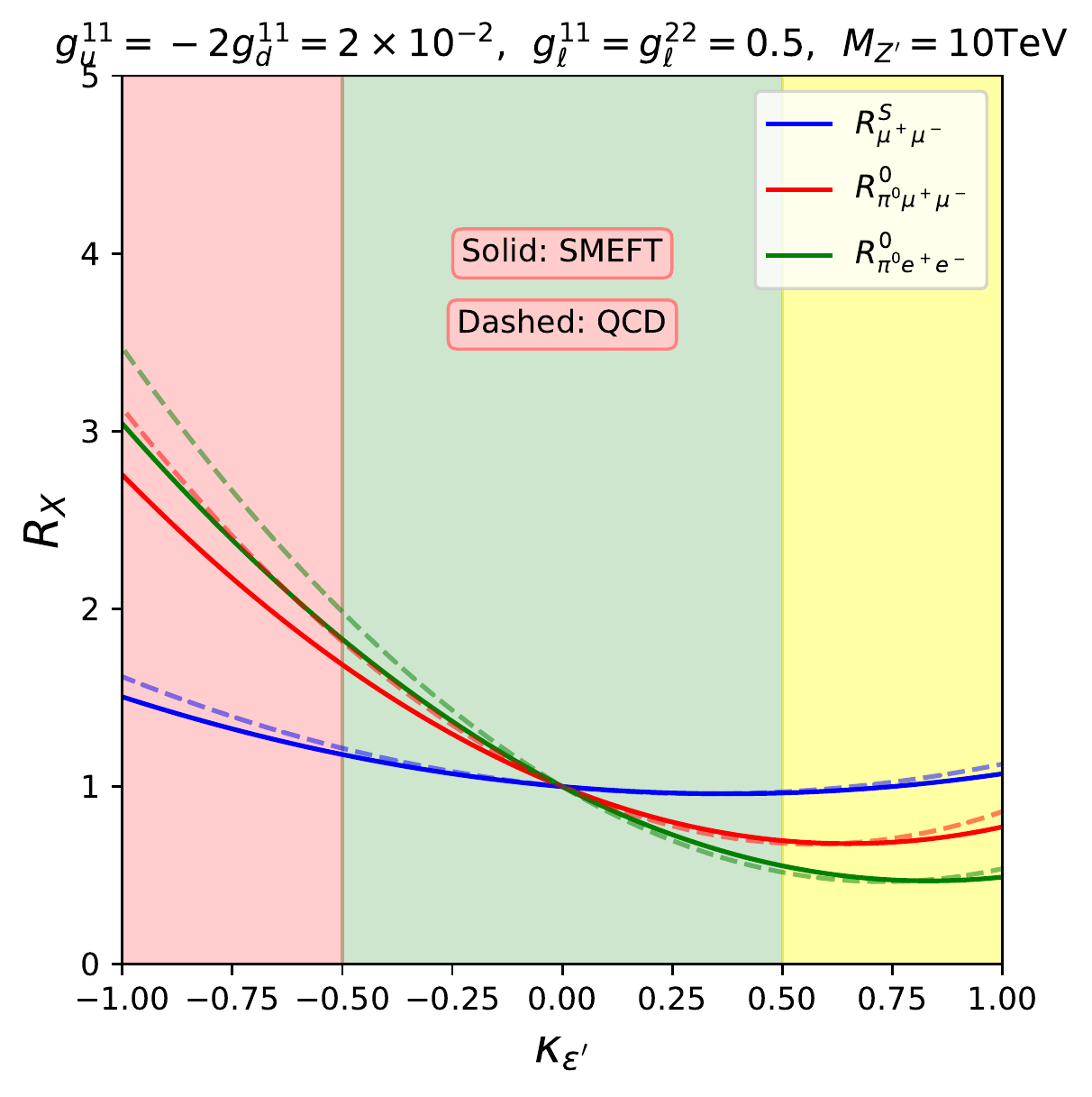}
\captionsetup{width=0.9\textwidth}
\caption{LH-EWP scenario for a $Z'$ of $3\tev$ (left panel) and $10\tev$ (right panel). The predictions for the ratios of the decays $K_S \to \mu^+ \mu^-$, $K_L \to \pi^0 \mu^+ \mu^-$ and $K_L \to \pi^0 e^+ e^-$ defined in eq.~\eqref{eq:kappas} are plotted against $\kappa_{\varepsilon'}$ . The yellow, green and red shades correspond to
the $\kappa_{\varepsilon^\prime}$ scenarios A, B and C as defined in \eqref{eq:epsp_cases}.}
\label{fig:LHS_kappas2}
\end{figure}

\noindent

This implies a strong correlation between $K^+\to\pi^+\nu\bar\nu$ and $K_L\to\pi\nu\bar\nu$ on the MB-branch,
 so that they can be enhanced or suppressed only simultaneously as shown by
 the orange colour in this figure.
{Out of the three $\kepe$ scenarios A, B and C,
which are defined in \eqref{eq:epsp_cases}, in scenario A,} large departures
  from SM expectations for $\klpn$ are possible.
  Similarly, in Fig.~\ref{fig:LHS_kp_k0} (right), the correlations between the ratio for the
  decay $K_L\to\pi\nu\bar\nu$ and  the ones for $K \to\pi \ell^+ \ell^-$ and $K_S \to \mu^+ \mu^-$ are shown. The upper range for $R_{\nu\bar\nu}^0$ corresponds roughly
  to the GN bound. If the values from KOTO given in (\ref{KOTO})
  will be confirmed in the future, large departures from the SM predictions for
  the three rare decays are to be expected. Also the $\kpn$ branching ratio could
  be enhanced. Fig.~\ref{fig:LHS_kp_k0} (right) admits two solutions for each decay, corresponding to different values of $\kappa_{\varepsilon'}$. The upper branch results from positive
values for ${\rm Im}(g_q^{21})$ and the lower one from negative ones, since positive (negative)
values of ${\rm Im} (g_q^{21})$ enhance (reduce) the corresponding ratios.

\noindent
In Fig.~\ref{fig:LHS_kappas1} we show the results for the first three different ratios
defined in (\ref{eq:kappas}) as functions of $\kappa_{\varepsilon'}$ for a $Z'$ of $3\tev$ and $10\tev$ respectively. For the running
below the EW scale we use the complete 1-loop QCD and QED running \cite{Aebischer:2017gaw,Jenkins:2017dyc}
and above the EW scale the full SMEFT RGEs for the solid and only QCD for the dashed
  lines {are} used. Clearly, the running is dominated by QCD effects.} For
  $3\tev$  both
  $K\rightarrow \pi \nu \bar \nu$ branching ratios are enhanced  over their
  SM values, except for a small region around  $\kepe\approx 0$. For
  $10\tev$, significant BSM effects are only observed for  $\kepe\ge 0.5$.
 $\Delta M_K$ is visibly suppressed for sufficiently large $\kepe$.
The choice of very small values of $g_{u,d}^{11}$
of $\ord(10^{-2})$ is implied, as noticed already in \cite{Buras:2015jaq}, by the desire to
suppress  $\Delta M_K$ in the presence of NP contributions to $\epe$ in
  the EWP sector. For  $g_{u,d}^{11}$ of
$\ord(1)$ considered in the latter paper, $\Delta M_K$ is enhanced by BSM rather than suppressed which is disfavoured by the present LQCD data.
\noindent
In Fig.~\ref{fig:LHS_kappas2} we show predictions for the remaining ratios given
in \eqref{eq:kappas}, where we allow for additional couplings to left-handed electrons ($g_{\ell}^{11}$).
We observe that for a lighter $Z^\prime$ an enhancement for $R^0_{\pi^0\mu\mu}$ and $R^0_{\pi^0ee}$
 processes is predicted for {negative values of $\kepe$, while
for its positive values both suppression as well as {enhancement} are possible.
 On the other hand for heavier $Z^\prime$
 these decay modes are  suppressed (enhanced) for positive (negative) values of $\kepe$.
The ratio $R_{\mu\mu}^S$  is always enhanced.
The difference between solid and dashed lines is mainly due to QED RG effects on $\kepe$, generated by
semi-leptonic operators.

 In Fig.~\ref{fig:LHS_kappas4} the correlations between $\kepe$ and
  $R_{\Delta M_K}$ and between the ratios for $\kpn$ and $\klpn$ and $R_{\Delta M_K}$ are shown. As expected, $\kepe$ and $\klpn$ are much more sensitive to
  variations of $R_{\Delta M_K}$ than it is the case of $\kpn$.

\noindent
In Fig.~\ref{fig:LHS_kappas3} the ratios of Fig.~\ref{fig:LHS_kappas2} are shown this time
as a functions}  of $R_{\Delta M_K}$ for a $Z'$ of $3\tev$ and $10\tev$.
 A large enhancement for all processes is possible for
{both light as well as heavy $Z^\prime$},
while suppressing $\Delta M_K$.
The sign of the quark coupling $g_q^{21}$ can be fixed by $\kepe$ if
the {signs of the} diagonal quark couplings are known. Similarly the leptonic couplings
can be either positive or negative and are not determined by the conditions imposed.
The two branches in this figure correspond to different signs of the coupling $g_q^{21}$. In any case the hinted $\Delta M_K$ anomaly has significant
  impact on all branching ratios.
\begin{figure}[htb]
\centering
\includegraphics[width=0.4\textwidth]{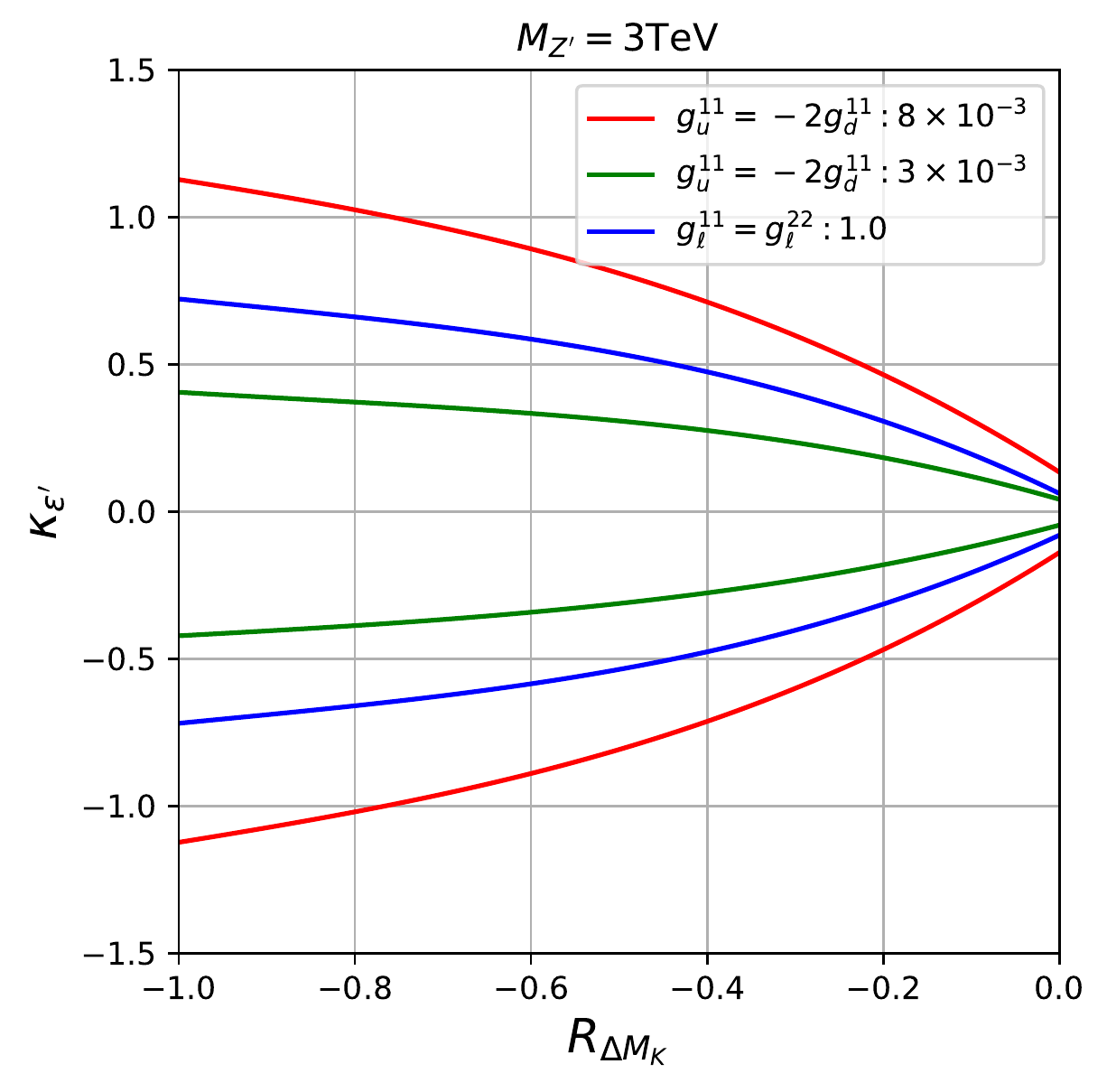}
\includegraphics[width=0.4\textwidth]{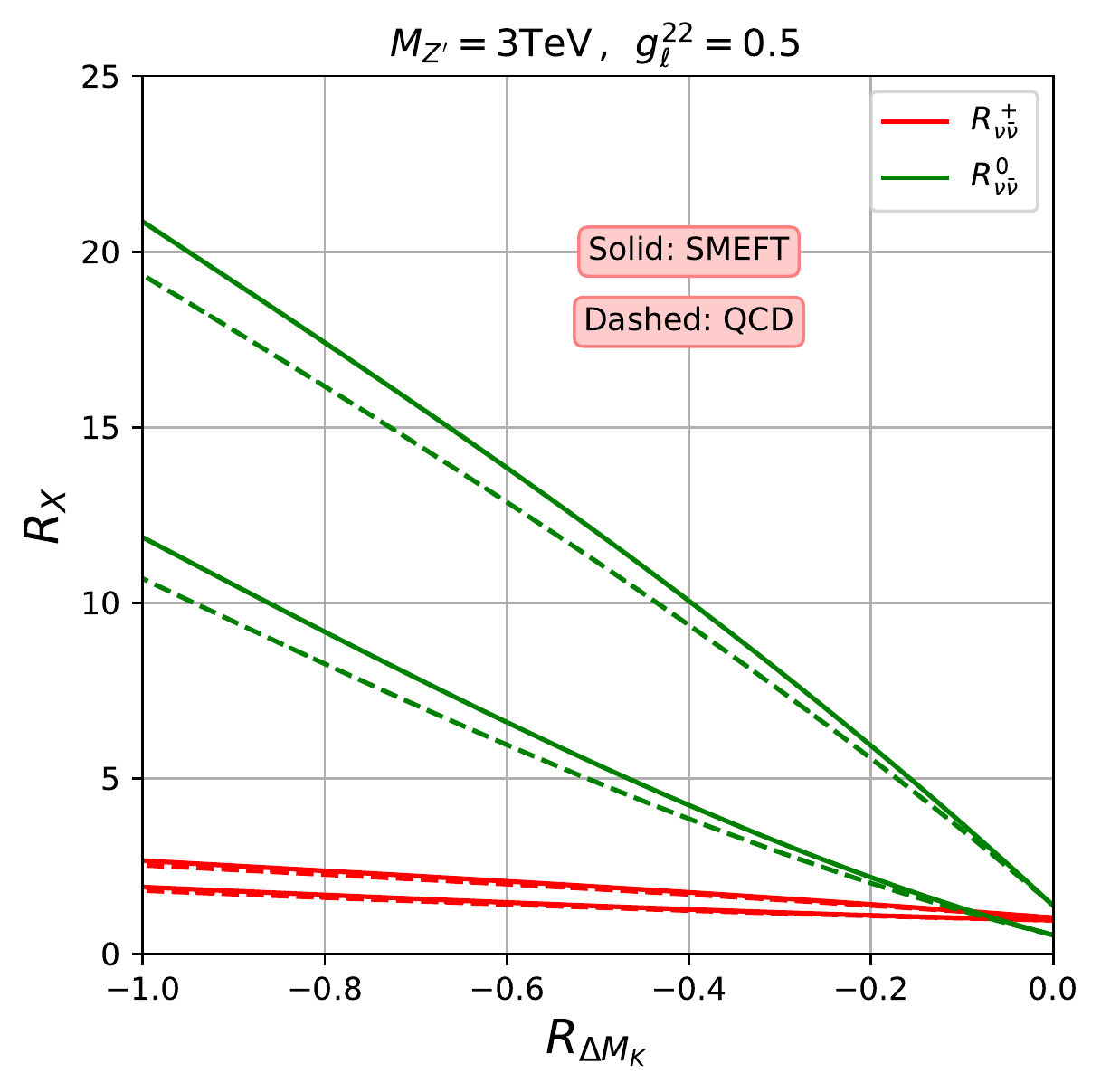}
\captionsetup{width=0.9\textwidth}
\caption{{LH-EWP scenario for a $Z'$ of $3\tev$.
The $\kappa_{\varepsilon'}$ and ratios for the process $K^+\to\pi^+\nu\bar\nu$,
$K_L\to\pi\nu\bar\nu$ (right) defined in \eqref{eq:kappas}} are plotted against $R_{\Delta M_K}$.}
\label{fig:LHS_kappas4}
\end{figure}
\begin{figure}[htb]
\centering
\includegraphics[width=0.4\textwidth]{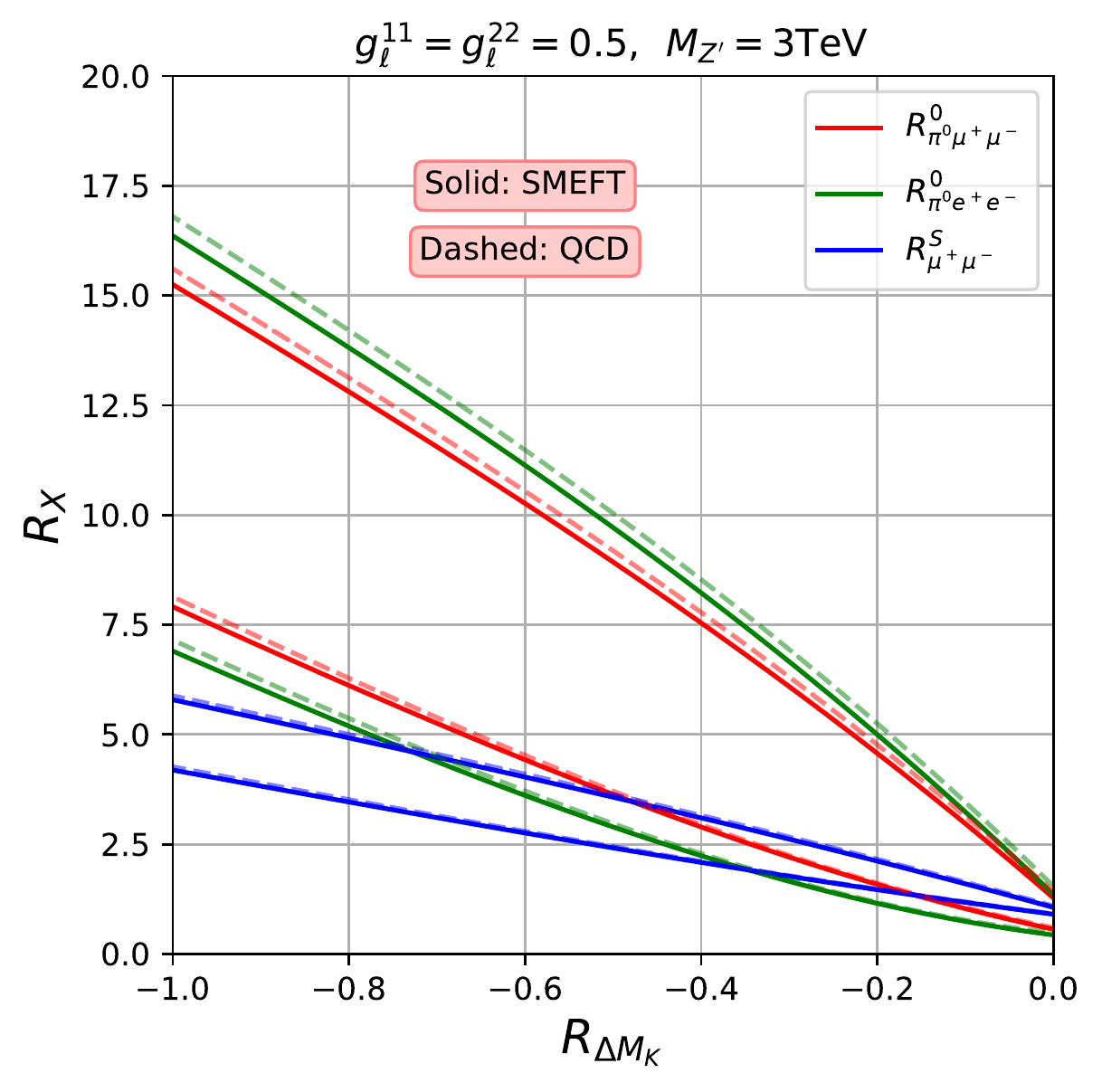}
\includegraphics[width=0.4\textwidth]{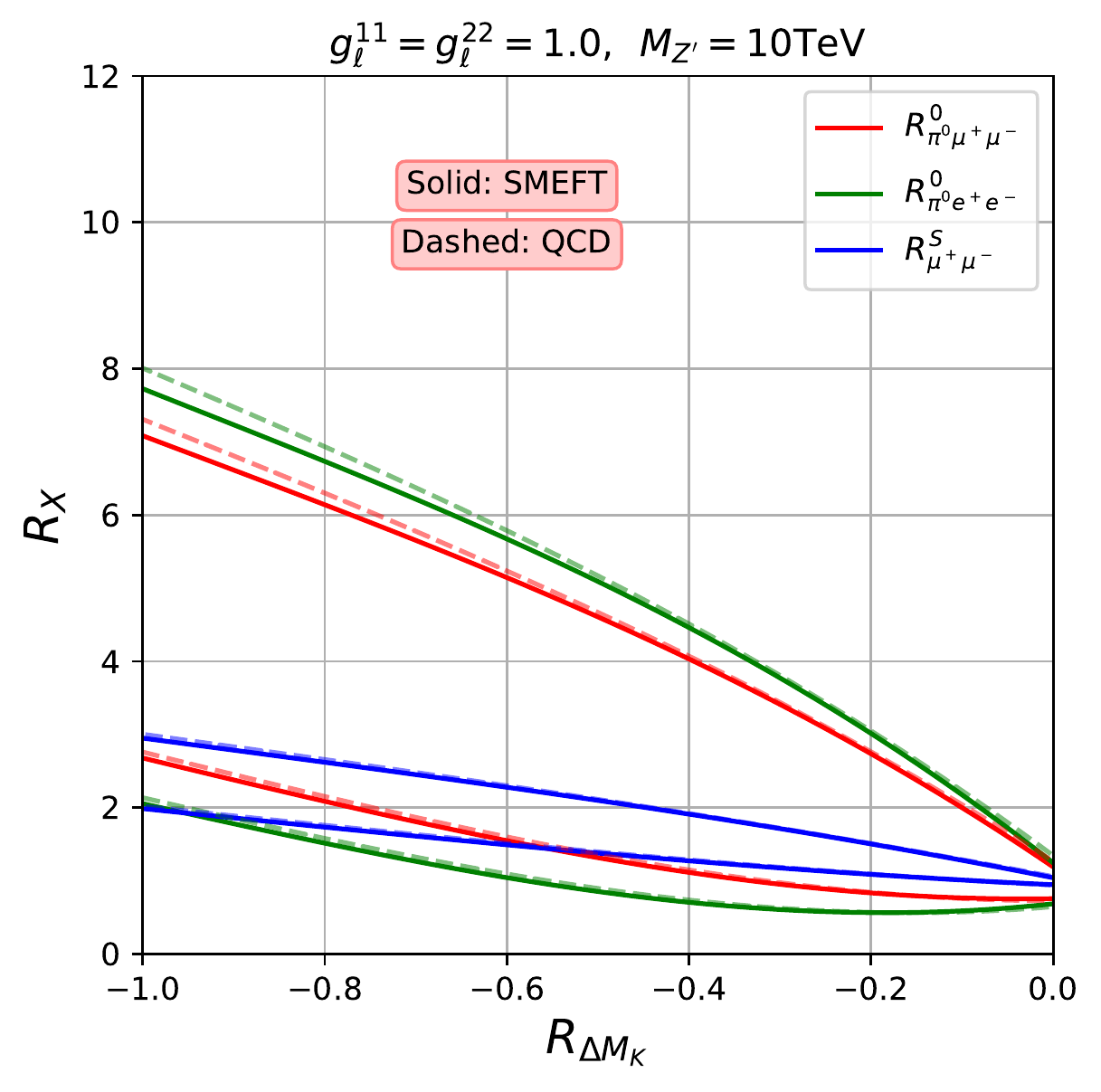}
\captionsetup{width=0.9\textwidth}
\caption{LH-EWP scenario for a $Z'$ of $3\tev$ (left panel) and $10\tev$ (right panel).
The predictions for the ratios of the decays
$K_S \to \mu^+ \mu^-$, $K_L \to \pi^0 \mu^+ \mu^-$ and $K_L \to \pi^0 e^+ e^-$ defined
in \eqref{eq:kappas} are plotted against $R_{\Delta M_K}$.}
\label{fig:LHS_kappas3}
\end{figure}
\subsection{QCD Penguin Scenario: Left- and Right-Handed}\label{rerotation}
Next we describe the effects related to the required basis rerotation at the
electroweak scale, as described in the last point of
Sec. \ref{sec:basisrot}. This has important phenomenological consequences in any scenario,
as for example in the QCD penguin (QCDP) scenario, in which a sizable imaginary coupling is {present in scenarios A and C for $\kepe$}. The LH-QCDP scenario is defined as follows:

\begin{equation}
g_q^{21}\neq 0\,,\qquad  g_u^{11}=g_d^{11} \qquad \text{(LH-QCDP scenario)}\,.
\end{equation}

\noindent
Starting with a set of non-zero Wilson coefficients in the down-basis at the high scale $\Lambda$ we evolve them to the EW scale. Along with
the Wilson coefficients we  also need to evolve the SM parameters including the mass (or Yukawa) matrices  as discussed in Section~\ref{sec:basisrot}. But the running of the mass matrices is flavour dependent\cite{Jenkins:2013wua}, and consequently  after the evolution the mass matrices are not guaranteed to remain in the original basis that we started with. As a result, we need to rotate the mass matrices and hence the Wilson coefficients to adhere to our choice  of the down-basis\cite{Aebischer:2018bkb}.  This issue is discussed in generality in a recent paper \cite{Aebischer:2020lsx} but here we confine our discussion focusing on QCDP.

We illustrate this effect and its phenomenological consequences with a concrete example by considering the {LH-QCDP} scenario studied in the case of significant BSM contributions to  $\epe$ in \cite{Buras:2015jaq}, but now in contrast to that paper including RG SMEFT effects.
Considering the LHS, at the high scale $\Lambda$ the operators $\op[(1)]{qu}{2111}$ and $\op[(1)]{qd}{2111}$ are generated. {They are then evolved down to the EW scale. But} the simultaneous evolution of
 the mass matrices  generates off-diagonal entries in the down-quark Yukawa matrix  $Y_d$ at the EW scale. This is due to the fact that the {running} of $Y_d$ is proportional to the up-quark Yukawa matrix $Y_u$, which is non-diagonal in the down-basis\cite{Machacek:1983fi}. Indeed, we have
\begin{equation}
16\pi^2\frac{d Y_d}{d \ln{\mu}} \simeq -\frac{3}{2}  (Y_u{Y_u}^{\dagger})  Y_d  + ...\,.
\end{equation}
To revert to the down-type basis, a rotation of the operators is necessary, as already explained in Section~\ref{sec:basisrot}. Applying this {back-rotation} to the Wilson coefficients generates $\wc[(1)]{qd}{2121}$ at the EW scale in the down-basis as:

\begin{eqnarray}\label{eq:rerot}
  \wc[(1)]{qd}{2121} &=& (U^\dag_{d_L})_{22}(U_{d_L})_{11}(U^\dag_{d_R})_{21}(U_{d_R})_{11}\wc[(1)]{qd^\prime}{2111}+
  ...\,,
\end{eqnarray}
where $\wc[(1)]{qd^\prime}{2111}$ denotes the Wilson coefficient in the RGE basis and the rotation matrices $U_{d_L},\,U_{d_R}$ satisfy the following equation:
\begin{equation}
	M_d(\mu_{\rm EW}) = U_{d_L}^\dagger M_d^\prime(\mu_{\rm EW}) U_{d_R}\,.
\end{equation}
Here the (non-diagonal) down-quark mass matrix $M_d^\prime$ at the EW scale is obtained by evolving $Y_d$ from the high scale $\Lambda$ { down to $\mu_{\rm EW}$. In the LL approximation we have:
\begin{equation}
	M_d^\prime( \mu_{\rm EW}) = M_d(\Lambda) + \frac{v}{\sqrt 2} \frac{\beta_{Y_d}(\Lambda)}{16 \pi^2} \ln{ \left (\frac{\mu_{\rm EW}}{\Lambda} \right)}\,.
\end{equation}
However, the Wilson coefficient $\wc[(1)]{qd}{2121}$} is strongly constrained
 by  $\varepsilon_K$ due to the large hadronic matrix element multiplying it.
\begin{figure}[htb]
\begin{center}
\includegraphics[width=0.5\textwidth]{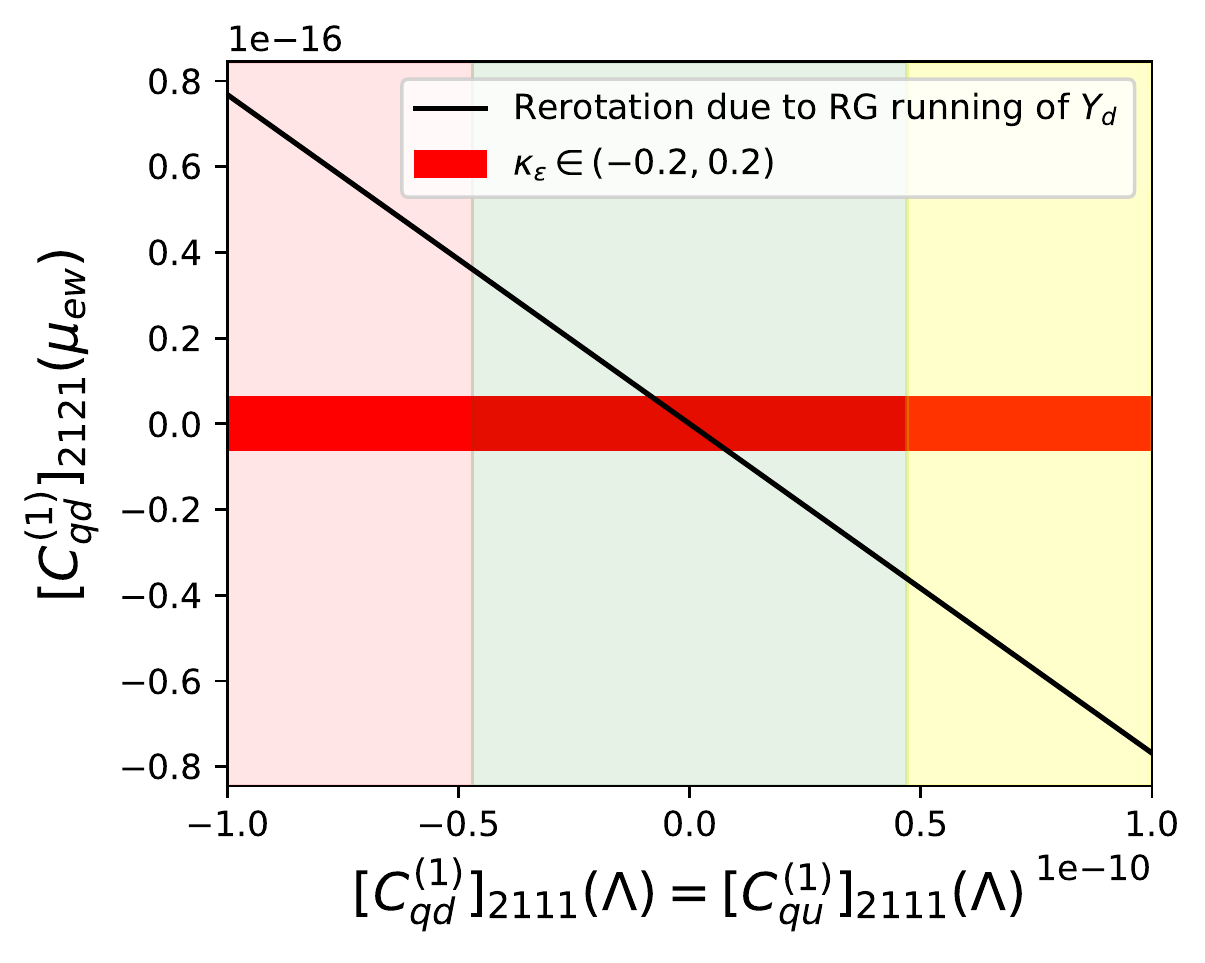}
\captionsetup{width=0.9\textwidth}
\caption{{LH-QCDP} scenario, where the operator $\op[(1)]{qd}{2121}$ is
        generated after RGE running of $Y_d$ and { back-rotation} to the down-basis at the EW scale.
        The allowed regions for the Wilson coefficients are in red for $\varepsilon_K$,
{and vertical bands represent the three $\epe$ scenarios. } }
\label{fig:Kumareffect}
\end{center}
\end{figure}
This scenario is illustrated in Fig.~\ref{fig:Kumareffect}, where at the high scale we vary the input
values of the Wilson coefficients $\wc[(1)]{qd}{2111}$ and $\wc[(1)]{qu}{2111}$ as shown on the x-axis. On the y-axis we show the output value of the Wilson coefficient  $\wc[(1)]{qd}{2121}$ at the EW scale which is generated through the back-rotation of \eqref{eq:rerot}. However, this LR operator gives a large
contribution to $\varepsilon_K$\cite{Buras:2015jaq}
\begin{equation}
\kappa_{\varepsilon} \simeq 3.1 \times 10^{16} \times  {\rm Im}
\left (\wc[(1)]{qd}{2121}(\mu_{ew}) {\rm GeV^{-2}} \right ).
\end{equation}

The  allowed values for the Wilson coefficients of the three mentioned operators are shown in the {red region, given the constraints from $\varepsilon_K$}. This shows that in the LHS with QCDP dominance, significant BSM
  contributions to $\epe$ imply a large contribution to $\varepsilon_K$ inevitably generated by
the running of Yukawas and subsequent {back-rotation} of the Wilson coefficients at the EW scale.
Consequently, the QCDP scenario for $\epe$, considered in \cite{Buras:2015jaq} is ruled out,
{since in this case} significant BSM contributions to $\epe$ would be required to fit the data.
 Similar comments
  apply to the RHS scenario defined by
\begin{equation}
g_q^{11}\neq 0\,,\qquad  g_d^{21} \ne 0\,. \qquad \text{(RH-QCDP scenario)}
\end{equation}
In this case only the QCDP scenario can be constructed. Due to
   $SU(2)_L$ gauge invariance the coefficient of the so-called $Q_8^\prime$ operator,
which otherwise would give a leading contribution to $\epe$, vanishes.

On the other hand,  in the case of the EWP dominance i.e $\wc[(1)]{qd}{2111} = -\frac{1}{2}\wc[(1)]{qu}{2111}$,  also considered in \cite{Buras:2015jaq}, this effect is negligible. This is simply because in this case a much smaller value of $\wc[(1)]{qd}{2111}$ is needed to enhance
sufficiently $\epe$.

\subsection{Left-Right Scenario}\label{LRSCENARIO}
We have just seen that in the LHS there was a very strong correlation between
$\kpn$ and $\klpn$ branching ratios on the {MB-branch}. As explained
in \cite{Blanke:2009pq} this {strict} correlation originates in the same complex phase
present in NP contributions to  $\varepsilon_K$  and rare Kaon decays in question
provided NP contributions to $\varepsilon_K$ are small.  This is in fact
evident in our case because the same $Z^\prime \bar s d$ coupling enters both
$\kpn$ and $\klpn$ and $\varepsilon_K$.

Now,
\be\label{EPSILONNEW}
(\varepsilon_K)_\text{BSM}\propto\left[(\RE(g_{sd})(\IM(g_{sd})\right],
\ee
and to make sure that this contribution is small either $\RE(g_{sd})$ or
$\IM(g_{sd})$ must be small. If $\IM(g_{sd})$ is small the horizontal
line in Fig.~\ref{fig:LHS_kp_k0} results with NP basically only in $\kpn$. If  $\RE(g_{sd})$
is small then {there} are NP contributions to both $\klpn$ and $\kpn$ correlated
on the MB-branch. In our case this second solution is chosen by the desire
to explain the $\Delta M_K$ anomaly. However, such a correlation
precludes
the pattern of simultaneously enhancing $\klpn$ and suppressing $\kpn$
possibly hinted by {the} NA62 and KOTO results.

It is known from various studies that such a pattern can be obtained through
the introduction of new operators and the most effective in this respect are
scenarios in which both left-handed and right-handed flavour-violating NP
couplings are present, breaking {the} correlation between $K^0-\bar K^0$ mixing
and rare Kaon decays {and} thereby eliminating the  impact of {the}
$\varepsilon_K$ constraint on rare Kaon decays.
The presence of left-right operators requires some fine-tuning of {the} parameters
in order to satisfy the $\varepsilon_K$ constraint but such operators
do not contribute to rare decays and the presence of new parameters does
not affect directly these decays. Examples of such scenarios are
$Z^\prime$ models with LH and RH couplings considered in \cite{Buras:2014zga} and the earlier
studies in the context of the general MSSM
\cite{Nir:1997tf,Buras:1997ij,Buras:1999da,Buras:2004qb,Isidori:2006qy} and Randall-Sundrum models \cite{Blanke:2008yr,Bauer:2009cf}. See in particular Fig.~6 in \cite{Blanke:2008yr} and Fig~7 in  \cite{Buras:2014zga}.
Needless to say also the correlations between NP contributions to $\Delta M_K$
and rare decays are diluted, although the necessity of non-vanishing complex
couplings required by the hinted $\Delta M_K$ anomaly will certainly have some impact on rare Kaon decays.

The Left-Right (LR) scenario at $3\tev$ is defined by
\begin{equation}
g_q^{21}, g_d^{21} \neq 0\,, \qquad  g_u^{11}=-2g_d^{11}\,, \qquad \text{(LR-EWP scenario)}
\end{equation}
which is equivalent to the LH-EWP scenario without imposing $\Delta F=2$
constraints\cite{Buras:2014zga}.

\begin{figure}[htb]
\centering
\includegraphics[width=0.5\textwidth]{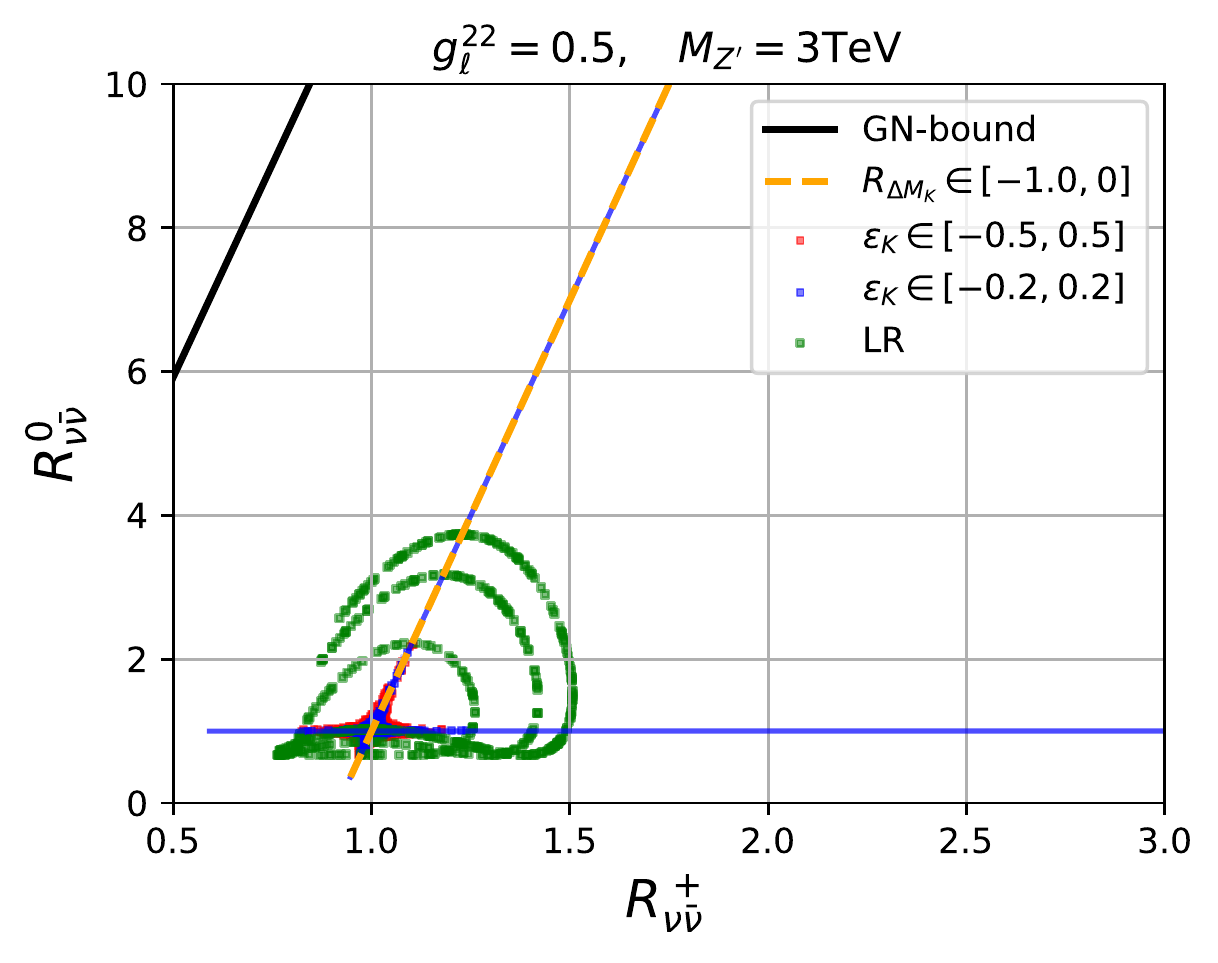}
\captionsetup{width=0.9\textwidth}
	\caption{{The ratios for $K^+\to\pi^+\nu\bar\nu$ and $K_L\to\pi\nu\bar\nu$ defined in
\eqref{eq:kappas} are plotted. The LR scenario shown in green and LH-EWP scenario in blue and red with
 $\varepsilon_K \in [-0.2,0.2]$ and $[-0.5,0.5]$
respectively for a $Z'$ of $3\tev$. The orange line also satisfies $R_{\Delta M_K} \in [-1.0,0]$. The GN bound is shown in black.    }  }
\label{fig:epsK}
\end{figure}

In Fig.~\ref{fig:epsK} correlations between ratios for $\klpn$ and $\kpn$ as in \eqref{eq:kappas}
are considered. Clearly no strong correlation is observed when both LH and RH {couplings}
are allowed as shown in the green region. Similarly, the strong correlation between $\kpn$ and $\kepe$ observed
in the LH-EWP scenario is absent in the LR scenario because $R^+_{\nu\bar\nu}$
also depends on the real part, which is not fixed through $\epe$.

Imposing however the constraint from $\varepsilon_K$ and therefore studying
a LH-EWP scenario limits the allowed parameter space drastically. Furthermore, as shown in
Sec.~\ref{subsec:LHSEWP} out of the two branches in the $R^+_{\nu\bar\nu}$-$R^0_{\nu\bar\nu}$ plane
 allowed by $\varepsilon_K$, the horizontal branch shown in blue
is disfavored by {the} requirement of suppression of $\Delta M_K$.
In the red area we show the allowed region for the LH-EWP scenario
with $\varepsilon_K \in [-0.5,0.5]$.

Importantly, as evident from Fig.~\ref{fig:epsK}, the simultaneous enhancement of $\klpn$ and
suppression of $\kpn$ branching ratios is only possible in the presence
  of both LH and RH flavour-violating couplings. Also, the observables $R_{\nu\bar \nu}^0$ and $\kepe$ only depend on the imaginary part
of the flavour violating coupling. Therefore they are strongly correlated in the LR as well as
in the LHS scenario.

This agrees
with the findings in \cite{Buras:2015yca}, in which only QCD has been considered.
The correlation between $R^+_{\nu\bar\nu}$ and $R^0_{\nu\bar\nu}$ in this setup is
therefore invariant under Yukawa running effects.

\section{$Z$ Contributions: Numerics}\label{sec:4}
\begin{table}[tbp]
\centering
\renewcommand{\arraystretch}{1.5}
\begin{tabular}{cc}
\toprule
& $\psi^2 H^2 D$
\\
\hline
$\Op[(1)]{H\ell}$
& $(H^\dagger i \overleftrightarrow{D}_{\!\!\mu} H) (\bar{\ell}^i \gamma^\mu \ell^j)$
\\
$\Op[(3)]{H\ell}$
& $(H^\dagger i \overleftrightarrow{D}^I_{\!\!\mu} H) (\bar{\ell}^i \tau^I \gamma^\mu \ell^j)$
\\
$\Op{He}$
& $(H^\dagger i \overleftrightarrow{D}_{\!\!\mu} H) (\bar{e}^i \gamma^\mu e^j)$
\\
$\Op[(1)]{Hq}$
& $(H^\dagger i \overleftrightarrow{D}_{\!\!\mu} H) (\bar{q}^i \gamma^\mu q^j)$
\\
$\Op[(3)]{Hq}$
& $(H^\dagger i \overleftrightarrow{D}^I_{\!\!\mu} H) (\bar{q}^i \tau^I \gamma^\mu q^j)$
\\
$\Op{Hu}$
& $(H^\dagger i \overleftrightarrow{D}_{\!\!\mu} H) (\bar{u}^i \gamma^\mu u^j)$
\\
$\Op{Hd}$
& $(H^\dagger i \overleftrightarrow{D}_{\!\!\mu} H) (\bar{d}^i \gamma^\mu d^j)$
\\
$\Op{Hud}$
& $(\widetilde{H}^\dagger i {D}_{\!\mu} H) (\bar{u}^i \gamma^\mu d^j)$
\\
\bottomrule
\end{tabular}
\renewcommand{\arraystretch}{1.0}
\caption{Dimension-six $\psi^2 H^2 D$ operators in SMEFT.}
  \label{tab:psi2H2D}
\end{table}
\subsection{Preliminaries}
{In this section we consider flavour violating (FV) $Z$ couplings induced by FV $Z^\prime$
couplings through SMEFT RG running effects.} Let us consider the LL running from the BSM
scale $\Lambda$ to the EW scale $\mu_{\text{EW}}$. For the Wilson coefficients of the  $\psi^2H^2D$ operators
defined in Tab.~\ref{tab:psi2H2D} keeping only the top Yukawa coupling $y_t$ and neglecting
the terms of $\mathcal{O}(V_{ts}^2)$ and $\mathcal{O}({V_{tb} V_{ts}})$  one finds \cite{Celis:2017hod, Jenkins:2013wua}
\begin{align}
\wc[(1)]{H q}{ij}(\mu_{\rm EW})&= \frac{y_t^2}{8\pi^2}
   \left ( \wc[(1)]{qq}{3ji3} ( \Lambda)+2N_c\wc[(1)]{qq}{33ij}(\Lambda)-N_c\wc[(1)]{qu}{ij33}(\Lambda) \right)
\ln{\left(\frac{\mu_{\text{EW}}}{\Lambda}\right)} \,, \label{eq:phiqrgeHq1}\\
   \wc[(3)]{H q}{ij}( \mu_{\rm EW})&= -\frac{y_t^2}{8\pi^2}
      \wc[(1)]{qq}{i33j}(\Lambda)\ln{\left(\frac{\mu_{\text{EW}}}{\Lambda}\right)}\,, \label{eq:phiqrgeHq3}\\
\wc[]{H d}{ij}(\mu_{\rm EW})&= \frac{N_cy_t^2}{8\pi^2}
\left( \wc[(1)]{qd}{33ij}(\Lambda)-\wc[(1)]{ud}{33ij}(\Lambda) \right)\ln{\left(\frac{\mu_{\text{EW}}}{\Lambda}\right)}\,, \label{eq:phiqrgeHd}\\
\wc[]{H u}{ij}(\mu_{\rm EW})&= \frac{y_t^2}{8\pi^2}
\left   (N_c\wc[(1)]{qu}{33ij}(\Lambda)-2N_c\wc[]{uu}{ij33}(\Lambda)-2\wc[]{uu}{i33j}(\Lambda) \right)\ln{\left(\frac{\mu_{\text{EW}}}{\Lambda}\right)}\,, \\
\wc[(1)]{H \ell}{ij}(\mu_{\rm EW})&= \frac{N_cy_t^2}{8\pi^2}
  \left ( \wc[(1)]{\ell q}{ij33}(\Lambda)-\wc[]{\ell u}{ij33}(\Lambda) \right ) \ln{\left(\frac{\mu_{\text{EW}}}{\Lambda}\right)}\,,  \label{eq:phirgeHl1}\\
\wc[]{H e}{ij}(\mu_{\rm EW})&= -\frac{N_cy_t^2}{8\pi^2}
\left( \wc[]{eu}{ij33}(\Lambda)-\wc[]{qe}{33ij}(\Lambda) \right ) \ln{\left(\frac{\mu_{\text{EW}}}{\Lambda}\right) }\,, \label{eq:phirgeHe}
\end{align}
whereas $\Op[(3)]{H \ell}$ and $\Op[]{H ud}$ are not generated in this approximation.
Yukawa running effects therefore generate modified $Z$-couplings to the SM fermions.

We can now express the usual FC quark couplings of the $Z$ in terms of
$\Wc[(1,3)]{Hq}$, $\Wc{Hu}$ and $\Wc{Hd}$. We have first
\begin{align}
  \label{eq:Zcouplings}
  \mathcal{L}_{\bar\psi\psi Z}^{\rm BSM} &
  = Z_{\mu} \sum_{\psi = u,d} \bar \psi_i \, \gamma^{\mu} \left(
        [\Delta_L^{\psi}(Z)]_{ij} \, P_L
  \,+\, [\Delta_R^{\psi}(Z)]_{ij} \, P_R \right) \psi_j \,,
\end{align}
with $\psi=u,d$ distinguishing between $up$- and $down$-quark couplings.
These complex-valued couplings are related to the SMEFT Wilson coefficients
through \cite{Bobeth:2017xry}
\begin{equation}
  \label{eq:Z-Deltas:dim-6-WC}
\begin{aligned}
  \phantom{x}[\Delta^u_L(Z)]_{ij} &
  = -\frac{g_Z}{2} v^2 \left[\Wc[(1)]{Hq} - \Wc[(3)]{Hq}\right]_{ij} , \qquad &
  [\Delta^u_R(Z)]_{ij} &
  = -\frac{g_Z}{2} v^2 \wc{Hu}{ij}\,,
\\
  [\Delta^d_L(Z)]_{ij} &
  = -\frac{g_Z}{2} v^2 \left[\Wc[(1)]{Hq} + \Wc[(3)]{Hq}\right]_{ij} , \qquad &
  [\Delta^d_R(Z)]_{ij} &
  = -\frac{g_Z}{2} v^2 \wc{Hd}{ij}\,,
\end{aligned}
\end{equation}
where  $g_Z=\sqrt{g_1^2+g_2^2}$ and $v=246\gev$ is the electroweak vacuum expectation value.

{In the $Z^\prime$ scenario considered here the $\psi^2H^2D$ operators are
  generated through RG effects and are smaller than in the case where these operators are already
  present at the high scale \cite{Bobeth:2017xry, Endo:2018gdn, deBlas:2017xtg}.}
 For the time being we assume that this is not the case here but we
  will comment briefly on their possible impact on our analysis below.

\boldmath
\subsection{Impact of RG-Induced $Z$ on LH-EWP Scenario}
\unboldmath
 In this subsection we study an explicit example of FV $Z$ couplings induced by FV $Z^\prime$
couplings through SMEFT RG running effects
and its effect on the ratios in \eqref{eq:kappas}.
For this purpose we assume two scenarios: In the first one only direct contributions
from a $Z^\prime$ are generated at the matching scale. This corresponds to the LH-EWP setup
in Subsection~\ref{subsec:LHSEWP}. In the second one we allow for additional non-zero
couplings to the third generation quarks.
The up-type quark coupling will then generate through
\eqref{eq:phiqrgeHq1}  modified $Z$-couplings,
which induce an additional effect compared to the $Z^\prime$-only case. We choose the various
couplings at the matching scale as follows:
{
\begin{eqnarray}
        Z^\prime    &:&  g_{q}^{21} \ne 0\, \,\   g_u^{11} = -2g_d^{11} \ne 0, \,\  g_{\ell}^{22} \ne 0, \\
        Z^\prime +Z &:& \,\ Z^\prime \,\ + \,\ g_u^{33} = -2g_d^{33} \ne 0.
\end{eqnarray}
In the $Z^\prime +Z$ case non-zero values of the couplings
$g_q^{21}$ and $g_u^{33}$ lead to the flavour violating coupling
of the $Z$-boson \eqref{eq:Z-Deltas:dim-6-WC}
\begin{equation}\label{eq:Z-rge}
[\Delta_L^d(Z)]_{21} = g_Z\frac{y_t^2 N_c}{16 \pi^2}    v^2\wc[(1)]{qu}{2133}
\ln{ \left(\frac{ \mu_{ \rm EW } } {\Lambda }
\right )}.
\end{equation}
Since the {usual} SM $Z$ couplings obey the relation
\begin{equation} \label{eq:Z-SM-quark}
        [\Delta_R^u(Z)]_{11} = - 2 [\Delta_R^d(Z)]_{11}\,,
\end{equation}
the operators $Q_7$ and $Q_8$ are generated through matching and QCD running, respectively.
The $Z$ contributions to $\epe$ generated from a $Z^\prime$ via RGE running are
therefore of the EWP type.
}

\begin{figure}[H]
\centering
\includegraphics[width=0.4\textwidth]{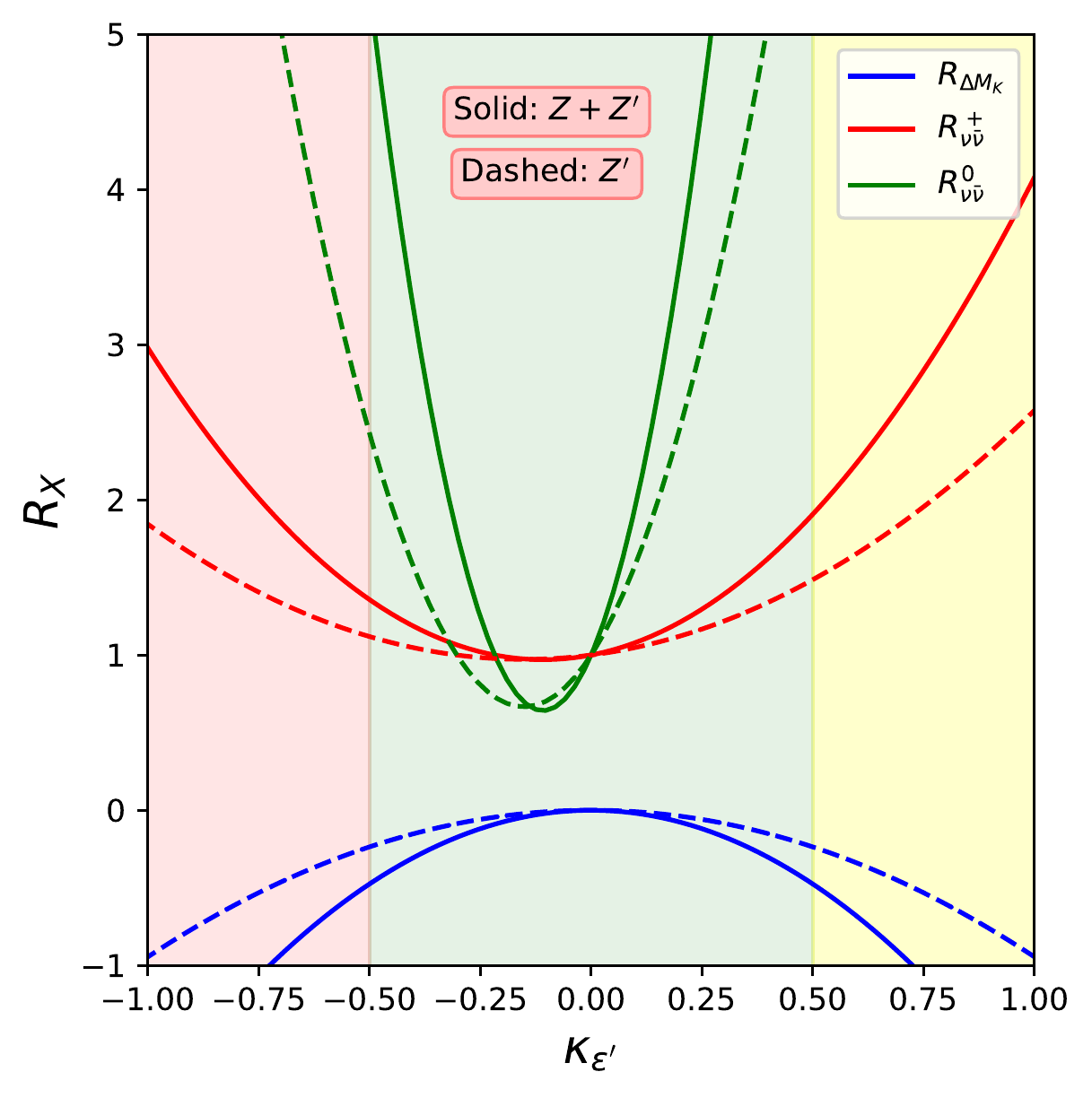}
\includegraphics[width=0.4\textwidth]{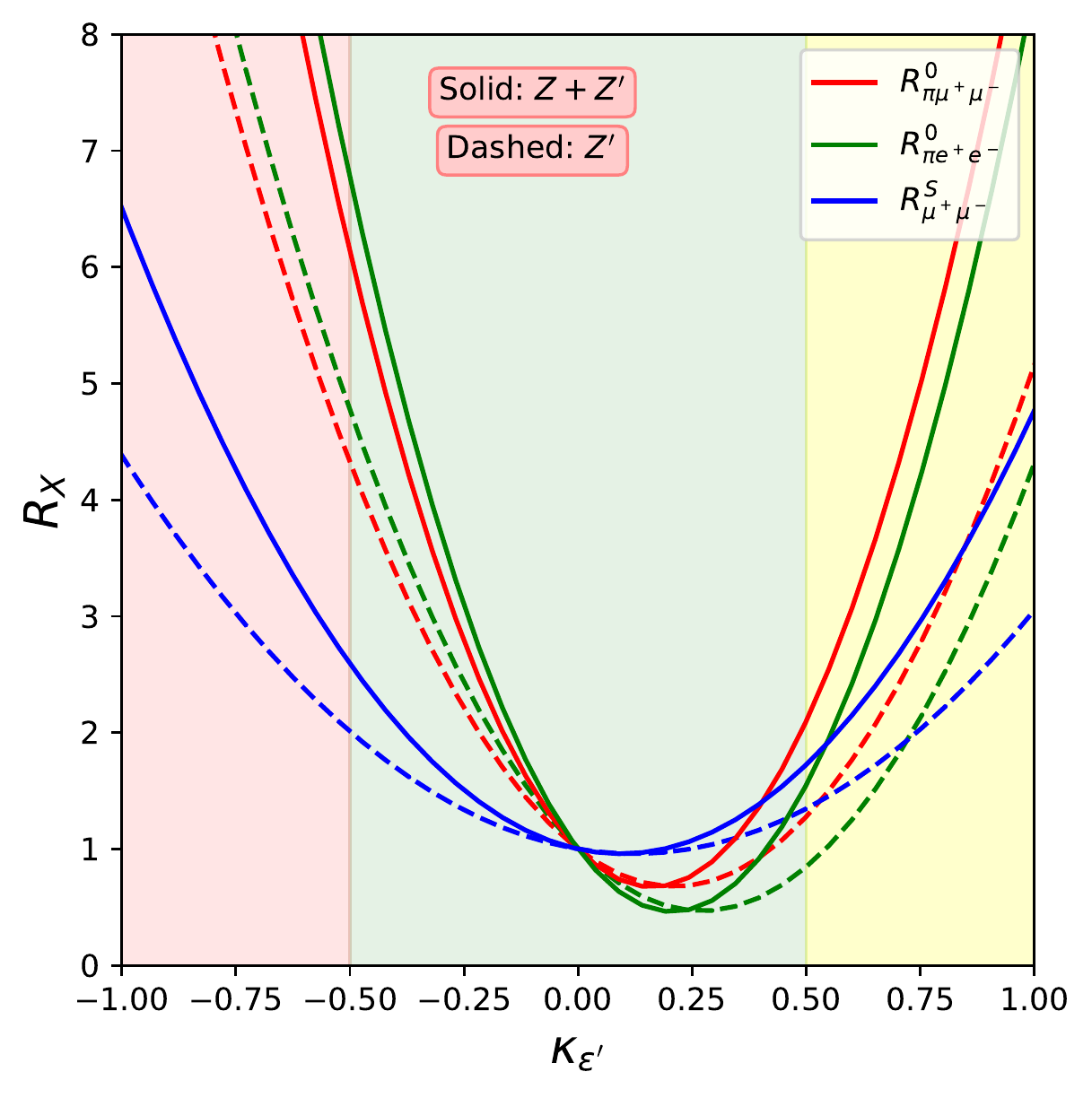}
\captionsetup{width=0.9\textwidth}
\caption{This figure shows how the $Z$-contributions to $\epe$ and other Kaon observables are generated from a $Z^\prime$
through RG running.}
\label{fig:epsp-phiq-RGE2}
\end{figure}

The results for the above two scenarios are shown
in Fig.~\ref{fig:epsp-phiq-RGE2}, where
 for a $Z^\prime$ of 3$\tev$ { the same values for the couplings
as in Fig.~\ref{fig:LHS_kappas1} are assumed. In addition we have}
\begin{equation}
 g_{u}^{33}=-2g_{d}^{33}=0.1\,,
\end{equation}
for the $Z^\prime +Z$ case.
 The dashed and solid lines correspond to
the $Z^\prime$ and $Z^\prime +Z$ case respectively. The additional contributions
due to the modified $Z$-couplings are destructive to $\kepe$ in this setup, { so that
a larger value of $g_q^{21}$ is needed in order to {obtain the} same value of $\kepe$ in the presence of
$Z$ contributions.
Therefore, for a given value of $\kepe$ the effect in semi-leptonic decays and $\Delta M_K$ is enhanced
as compared to the $Z^\prime$-solo scenario. }
By changing the sign of the third-generation couplings, a constructive effect can be achieved for $\kepe$.

In the left chart of Figure~\ref{fig:epsp-phiq-RGE2} $R^0_{\nu\bar\nu}$ and $R^+_{\nu\bar\nu}$
are enhanced whereas $\Delta M_K$ is suppressed.
The modified $Z$ contributions can have large influence on  $K_L\to\pi^0\nu\bar\nu$
which is less pronounced for $K^+\to\pi^+\nu\bar\nu$
 for  moderate values of $\epe$. The effect in $\Delta M_K$ is also less pronounced since the
modified $Z$ coupling enters quadratically. For the predictions of the (semi)-leptonic
decays in the right chart in Figure~\ref{fig:epsp-phiq-RGE2} the effect of the
 generated FV $Z$ coupling  is significant for larger absolute values of $\kepe$ and predicts
  enhancements of all considered ratios.

\boldmath
\subsection{$\epe$ and Rare decays from RG-Induced $Z$}
\begin{figure}[h]
\centering
\includegraphics[width=0.5\textwidth]{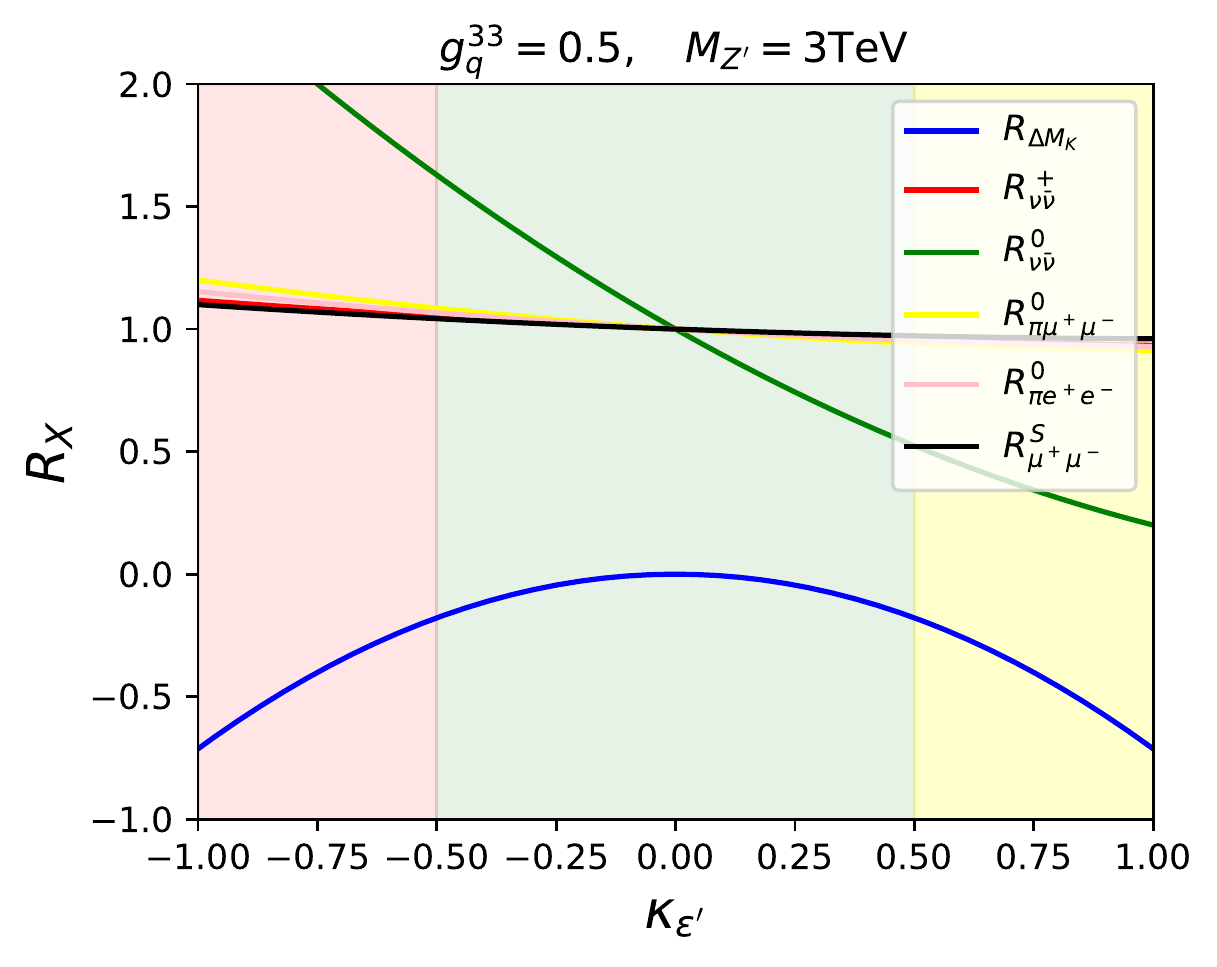}
\captionsetup{width=0.9\textwidth}
\caption{This figure shows how the $Z$-contributions to $\epe$ and other Kaon
observables are generated from {a} $Z^\prime$
 with purely left-handed quark couplings through RG running.}
\label{fig:epsp-phiq-RGE}
\end{figure}

\unboldmath
In our previous discussion we {found} that in order to
{have significant BSM contributions to $\epe$} within the EWP scenario
right-handed flavour diagonal couplings to the
first generation quarks are required.
However, in this subsection we show that one can also {get BSM contributions to
$\epe$ } even from purely left-handed $Z^\prime$ couplings. {This can happen}
through top-Yukawa RG running effects.
For this purpose we assume a scenario in which at the high scale the diagonal
couplings to the first generation quarks vanish and allow for a rather
 large third generation coupling, namely
\begin{equation}
g_q^{21}\neq 0\,,\quad	g_u^{11} = g_d^{11}= 0\,, \quad g_q^{33} = 0.5\,.
\end{equation}
This choice ensures vanishing of the direct $Z^\prime$  contribution to $\epe$ through {EWPs}.
In this setup the Wilson coefficient $\wc[(1)]{qq}{2133}$ is generated at the
BSM scale, which in turn generates $\wc[(1)]{Hq}{21}$ at the EW scale through top-Yukawa RGEs, as
 shown in \eqref{eq:phiqrgeHq1}. This leads to the flavour violating coupling
of the $Z$-boson \eqref{eq:Z-Deltas:dim-6-WC}
\begin{equation}\label{eq:anZ}
[\Delta_L^d(Z)]_{21} = - g_Z\frac{y_t^2 N_c}{8 \pi^2}    v^2\wc[(1)]{qq}{2133}
\ln{ \left(\frac{ \mu_{ \rm EW } } {\Lambda }
\right )}\,,
\end{equation}
which along with the usual SM $Z$ couplings \eqref{eq:Z-SM-quark},
generate the operators $Q_7$ and $Q_8$.
This effect is displayed in Figure \ref{fig:epsp-phiq-RGE}.
The different ratios of \eqref{eq:kappas} are shown as a function of $\kepe$.
A strong suppression of $\Delta M_K$ and correlation with $\epe$ is possible.
The large effect in $\Delta M_K$ is simply due to the sizable value of the flavour violating coupling
present at the BSM scale. {Except for $R_{\nu \bar \nu}^0$} all other ratios are almost
at their SM values and do not depend on $\kepe$.
In LHS or RHS $R^0_{\nu\bar\nu}$ goes down (up) with increased (decreased) $\kepe$ in $Z$-scenarios.
  This is because of special values of flavour diagonal $Zq\bar q$  couplings that equal the SM ones in this scenario. See the plots in \cite{Buras:2015yca,Buras:2015jaq}.

\noindent
In a similar fashion with different combinations of $Z^\prime$ couplings
at the NP scale the $Z$ couplings can be modified through other $\psi^2 H^2 D^2$
operators given in \eqref{eq:phiqrgeHq1}-\eqref{eq:phirgeHe}.

Finally, it should be emphasized following \cite{Bobeth:2017xry,Endo:2018gdn}  that
  $Z$ contributions to $\varepsilon_K$ and $\Delta M_K$ considered
  by us correspond really to dimension-eight operators, but the fact that
  the FV $Z$ couplings in rare decays and Wilson coefficients of these operators are the
  same implies correlations between $\Delta S=1$ and $\Delta S=2$ observables
  \cite{Buras:2012jb}. These correlations are strongly modified, even broken,
  in the presence of
  non-vanishing Wilson coefficients of $\psi^2H^2D$ operators already at the NP scale.
  Indeed, through top-Yukawa RG effects dimension-six operators contributing to
  $\varepsilon_K$ and $\Delta M_K$ are generated, implying in particular in the case of the
$\Op{Hd}$ operator
  strong  constraints
  on rare Kaon decays \cite{Bobeth:2016llm,Bobeth:2017xry,Endo:2018gdn}.

\section{Summary and Outlook}\label{sec:5}
The main goal of our paper was to confront $Z^\prime$ scenarios with the pattern of BSM contributions hinted by recent results on $\epe$, $\Delta M_K$, $\kpn$ and
$\klpn$ that appear to
\begin{itemize}
\item
 {allow significant positive or negative BSM contributions to $\epe$ relative to  its SM value,}
\item
  suppress the mass difference $\Delta M_K$ relative to the recent SM value
  obtained by the RBC-UKQCD collaboration,
\item
  suppress the branching ratio for $\kpn$ relative to the precise SM predictions as indicated by the
  recent result from the NA62 collaboration, although significant enhancements are still possible,
\item
  enhance the branching ratio for $\klpn$ relative to the precise SM prediction
  as hinted by the recent result from {the} KOTO collaboration.
\end{itemize}

Taking into account the constraints from $\varepsilon_K$ and $K_L\to\mu^+\mu^-$
we have calculated $\Delta M_K$ and the branching ratios for $\kpn$ and
$\klpn$ as functions of the parameter $\kepe$ introduced in
\cite{Buras:2015jaq} for the choices of $Z^\prime$ couplings to quarks and leptons that can reproduce the pattern of deviations from SM expectations summarized
above. For these choices of couplings we have calculated the
implications for $K_S\to \mu^+\mu^-$ and $K_L\to\pi^0\ell^+\ell^-$ again as functions of $\kepe$. Moreover, we have investigated correlations between all these
observables in various $Z^\prime$ scenarios.

While an analysis of this sort has been already presented in \cite{Buras:2015jaq}, prior to the last three hints for the pattern of BSM contributions,
and earlier analyses can be found in \cite{Buras:2012jb,Buras:2012dp}, this
is the first analysis of this set of observables to date that took into account
RG effects in the framework of the SMEFT, in particular the effects of top Yukawa couplings.

In this context we have  also
investigated for the first time whether  the presence of a heavy $Z^\prime$ with flavour violating couplings could generate through top Yukawa renormalization group effects  FCNCs mediated by the SM $Z$-boson.
Our results can be found in numerous plots. Here we want to list the most important lessons from our analysis.

{\bf Lesson 1:} While the correlation between the enhancement of $\epe$
with the suppression of $\Delta M_K$ has been already pointed out in the context
of the QCD penguin scenario for $\epe$ for flavour diagonal $Z^\prime$ couplings to quarks of $\ord(1)$ in \cite{Buras:2015jaq}, we find that the inclusion of RG top quark Yukawa effects rules out this scenario through the $\varepsilon_K$
constraint.

{\bf Lesson 2:} While, as noticed already in \cite{Buras:2015jaq}, the suppression of $\Delta M_K$ in the presence of the enhancement of $\epe$ could
in the EW penguin scenario be only obtained for
flavour diagonal $Z^\prime$ couplings to quarks of $\ord(10^{-2})$, a numerical  analysis of such a scenario has not been presented there. Our analysis demonstrates that the expectations from
\cite{Buras:2015jaq} are confirmed in the presence of the full RG SMEFT
analysis. In particular the $\varepsilon_K$ constraints are satisfied.

{\bf Lesson 3:} We point out that the present pattern of possible BSM effects
in $\kpn$ and $\klpn$ gives in the context of $Z^\prime$ models some indication for the presence of right-handed flavour violating currents at work. The confirmation of these findings requires in particular a much more accurate measurement of the $\kpn$ branching ratio by NA62. Otherwise a strong correlation between $\kpn$ and $\klpn$ branching
  ratios on the MB-branch is implied by the hinted $\Delta M_K$ anomaly. In this case if the
  large enhancement of $\klpn$ branching ratios signaled by the KOTO experiment
  is confirmed one day, also significant enhancement of the $\kpn$ branching
  ratio over its SM value is to be expected. As seen in Fig.~\ref{fig:LHS_kp_k0}, even larger departures
    from SM predictions should then be observed in $K \to\pi \ell^+ \ell^-$ and $K_S \to \mu^+ \mu^-$.

{\bf Lesson 4:} We have demonstrated that RG effects can in the presence
  of $Z^\prime$ contributions generate flavour-violating $Z$ contributions to $\epe$ and rare decays that   have significant impact on the phenomenology
  as shown in Fig.~\ref{fig:epsp-phiq-RGE2}. What we also find is that
  in the presence of $\ord(1)$ diagonal $Z^\prime$ top-quark couplings,
  the $(V-A)\times (V+A)$ EWP  operators can be generated solely
  through the
  RG induced flavour-violating $Z$ couplings. As shown in Fig.~\ref{fig:epsp-phiq-RGE} this effect is sufficiently strong to  provide significant  BSM contributions to $\epe$, if required, while
  simultaneously suppressing $\Delta M_K$.

{\bf Lesson 5:}  The impact of BSM effects on rare Kaon decays depends both on the scenarios  discussed and on the values of the couplings involved. With improved measurements it
  will be possibly to select the favorite scenarios. In this context
  the determination of the parameter $\kepe$ through improved LQCD
  calculations will be important because, as seen in several plots, some
  of the rare branching ratios depend sensitively on this parameter.

We are looking forward to experimental and theoretical developments in the coming years. Our plots will allow to monitor them and help to identify the successful $Z^\prime$ scenarios.

\section*{Acknowledgments}

J. A. acknowledges financial support from the Swiss National Science Foundation (Project No. P400P2\_183838). A.J.B acknowledges financial support from the Excellence Cluster ORIGINS,
funded by the Deutsche Forschungsgemeinschaft (DFG, German Research Foundation) under Germany's Excellence Strategy – EXC-2094 – 390783311.  J.K. acknowledges hospitality of Institue of Advanced Study (IAS) at TUM, Munich where this work was partially
completed. J.K. is supported by financial support from NSERC of Canada.
\appendix

\section{Hadronic matrix elements}
In this appendix we report the hadronic matrix elements we use for the numerics of $\epe$, which have been updated recently by the RBC-UKQCD collaboration
\cite{Abbott:2020hxn}. They are given in Tab.~\ref{tab:MEs}.
\begin{table}[H]
\centering
\renewcommand{\arraystretch}{1.2}
\begin{tabular}{crr}
  \toprule
  $Q_i$   & $\langle Q_i \rangle_0$                &  $\langle Q_i \rangle_2$
\\
\midrule
  $Q_3$     & $-0.075(57)(12)$  & $0$
\\
  $Q_4$     & $0.093(51)(15)$      & $0$
\\
  $Q_5$     & $-0.120(53)(19)$     & $0$
\\
  $Q_6$     & $-0.641(46)(101)$     & $0$
\\ \midrule
  $Q_7$     & $0.217(16)(34)$      & $0.0989(68)(30)$
\\
  $Q_8$     & $1.583(30)(249)$        & $0.683(19)(41)$
\\
  $Q_9$     & $-0.059(17)(9)$     & $0.0128(3)(8)$
\\
\bottomrule
\end{tabular}
\renewcommand{\arraystretch}{1.0}
\captionsetup{width=0.8\textwidth}
\caption{Numerical values of $K\to\pi\pi$ SM hadronic matrix elements used in our
  analysis.
}
\label{tab:MEs}
\end{table}

\addcontentsline{toc}{section}{References}

\small

\bibliographystyle{JHEP}
\bibliography{Bookallrefs}

\end{document}